\documentclass[12pt]{article}

\usepackage{newtxtext,newtxmath}

\usepackage{graphicx}

\usepackage[letterpaper,margin=1in]{geometry}

\renewenvironment{abstract}
	{\quotation}
	{\endquotation}

\date{}

\makeatletter
\renewcommand{\fnum@figure}{\textbf{Figure \thefigure}}
\renewcommand{\fnum@table}{\textbf{Table \thetable}}
\makeatother

\usepackage{scicite}

\usepackage{url}

\def\scititle{
	Starspots as the origin of ultrafast drifting radio bursts from an active M dwarf
}
\title{\bfseries \boldmath \scititle}

\author{
        Jiale Zhang$^{1,3,4}$,
        Hui Tian$^{1,2\ast}$,
        Stefano Bellotti$^{5,6}$,
        Tianqi Cang$^{7}$,\and
        Joseph R. Callingham$^{3,5}$,
        Harish K. Vedantham$^{3,4}$,
        Bin Chen$^{8}$,
        Sijie Yu$^{8}$,\and
        Philippe Zarka$^{9,10}$,
        Corentin K. Louis$^{9}$,
        Peng Jiang$^{11,12}$,
        Hongpeng Lu$^{1,13}$,\and
        Yang Gao$^{14}$,
        Jinghai Sun$^{11,12}$,
        Hengqian Gan$^{11,12}$,
        Hui Li$^{11,12}$,
        Chun Sun$^{11,12}$,\and
        Zheng Lei$^{11,12}$,
        Menglin Huang$^{11,12}$,\and
    \small$^{1}$School of Earth and Space Sciences, Peking University, Beijing 100871, China.\and
    \small$^{2}$State Key Laboratory of Solar Activity and Space Weather, National Space Science Center, \and \small Chinese Academy of Sciences, Beijing 100190, China.\and
    \small$^{3}$ASTRON, Netherlands Institute for Radio Astronomy, Oude Hoogeveensedĳk 4, Dwingeloo, 7991 PD, \and \small The Netherlands.\and
    \small$^{4}$Kapteyn Astronomical Institute, University of Groningen, P.O. Box 800, 9700 AV, Groningen, The Netherlands.\and
    \small$^{5}$Leiden Observatory, Leiden University, PO Box 9513, 2300 RA Leiden, The Netherlands.\and
    \small$^{6}$Institut de Recherche en Astrophysique et Plan\'etologie, Universit\'e de Toulouse, CNRS, IRAP/UMR 5277, \and \small 14 avenue Edouard Belin, F-31400, Toulouse, France.\and
    \small$^{7}$School of Physics and Astronomy, Beijing Normal University, Beijing 100875, China.\and
    \small$^{8}$Center for Solar-Terrestrial Research, New Jersey Institute of Technology, Newark, NJ 07102, USA.\and
    \small$^{9}$LIRA, Observatoire de Paris, Université PSL, Sorbonne Université, Université Paris Cité, \and \small CY Cergy Paris Université, CNRS, 92190 Meudon, France.\and
    \small$^{10}$ORN, Observatoire Radioastronomique de Nan\c{c}ay, Observatoire de Paris, CNRS, Univ. PSL, Univ. Orl\'{e}ans, \and \small F-18330 Nan\c{c}ay, France.\and
    \small$^{11}$National Astronomical Observatories, Chinese Academy of Sciences, Beijing 100012, China.\and
    \small$^{12}$CAS Key Laboratory of FAST, National Astronomical Observatories, Chinese Academy of Sciences, \and \small Beijing 100101, China.\and
    \small$^{13}$State Key Laboratory of Public Big Data and Guizhou Radio Astronomical Observatory, Guizhou University, \and \small Guiyang 550025,  China.\and
    \small$^{14}$School of Physics and Astronomy, Sun Yat-Sen University, Zhuhai 519082, Guangdong, China.\and
        \small$^\ast$Corresponding author.
    Email: huitian@pku.edu.cn\and
}    

\begin{document} 

\maketitle

\begin{abstract} \bfseries \boldmath
Detecting coherent radio bursts from nearby M dwarfs provides opportunities for exploring their magnetic activity and interaction with orbiting exoplanets. However, it remains uncertain if the emission is related to flare-like activity similar to the Sun or magnetospheric process akin to magnetized planets. Using observations (1.0 - 1.5 GHz) taken by the Five-hundred-meter Aperture Spherical radio Telescope, we found a type of millisecond-scale radio bursts with exceptionally high frequency drift rates ($\sim 8\;\rm{GHz\;s^{-1}}$) from an active M dwarf, AD Leo. The ultrafast drift rates point to a source region with a notably low magnetic scale height ($<0.15\; r_\star$, $r_\star$ as the stellar radius), a feature not expected in a commonly assumed dipole-like global field but highly possible in localized strong-field structures, i.e. starspots. Our findings suggest that a concentrated magnetic field above starspots could be responsible for some of the most intense radio bursts from M dwarfs, supporting a solar-like electron acceleration mechanism.

Teaser: Ultrafast drifting radio bursts from an M dwarf provide a method to probe stellar magnetic field structures.
\end{abstract}

\section*{Introduction}
\noindent
Radio observations are important to unveil many key physical processes on celestial objects in the solar system. Routine observations of solar radio emission are essential for monitoring eruptive magnetic activity such as flares and coronal mass ejections (CMEs)\cite{1985ARA&A..23..169D}. On magnetized planets like Earth and Jupiter, auroral radio emissions are observed in the polar regions and represent an important indicator of auroral activity\cite{1998JGR...10320159Z}. Recently, there has been a growing interest in detecting radio signals from nearby stellar systems, with the hope of exploring similar phenomena to those in the solar system. Despite these efforts, our knowledge of the radio emission from extrasolar systems remains at a rudimentary stage.

Among all the attempts, M dwarfs, the coolest and most common main-sequence stars in the Galaxy\cite{2006AJ....132.2360H}, have drawn great attention. Prior observations on M dwarfs have found a number of coherent radio bursts with strong circular polarization. They are often interpreted as electron cyclotron maser (ECM) emission\cite{2019ApJ...871..214V,2022ApJ...935...99B,2021A&A...648A..13C}, the same type of emission responsible for the radio aurorae observed in planetary magnetospheres\cite{1979ApJ...230..621W,2006A&ARv..13..229T}. This similarity implies that analogies could be made regarding the physical origins of radio emissions from M dwarfs and magnetized planets like Jupiter. For instance, an M dwarf may interact with a close-in planet akin to the Jupiter-Io system, producing Alfvén waves that accelerate the source electrons of the emission\cite{2007P&SS...55..598Z,2013A&A...552A.119S,2020NatAs...4..577V,2021NatAs...5.1233C,2023NatAs...7..569P}. A fast-rotating star may also involve co-rotation breakdown of plasma at a certain distance, which drives the auroral current system in the magnetosphere\cite{2012ApJ...760...59N,2015Natur.523..568H,2017MNRAS.470.4274T}. Alternatively, many active M dwarfs experience frequent magnetic activity in their coronae, acting as another source of energetic electrons. A recent solar radio observation reveals sustained ECM emission (for days) above a sunspot, which appears to be powered by recurring flares near the sunspot\cite{2024NatAs...8...50Y}. Similar inferences have also been drawn on some active M dwarfs based on recent stellar observations\cite{2019ApJ...871..214V,2020ApJ...905...23Z,2024A&A...686A..51M}.

Localization of the source of the radio bursts from stars is crucial for an accurate interpretation, as in the signature of the coronal activity or interaction with orbiting planets. However, it is challenging in practice to distinguish between emission coming from localized strong-field regions like our Sun and that from a global magnetic field like Jupiter, except for rare cases using very long baseline interferometry (VLBI)\cite{1998A&A...331..596B,2020A&A...641A..90C}. Our current understanding of the magnetic fields of M dwarfs primarily relies on two complementary approaches, Zeeman Doppler Imaging (ZDI) and Zeeman broadening (see ref. \cite{2021A&ARv..29....1K} for a review). The former applies tomographic techniques to reconstruct the globally large-scale surface magnetic field from circularly polarized spectra (Stokes V). This method tends to miss the small-scale structures with mixed polarities. In contrast, Zeeman broadening uses unpolarized spectra (Stokes I) to infer the total, unsigned magnetic field, and incorporates contributions from both large-scale and possible small-scale structures. It is recognized that the Stokes V measurements may miss up to 90\% of the total magnetic flux, meaning that a large portion of the magnetic energy is stored in small-scale structures on M dwarfs\cite{2009A&A...496..787R,2015ApJ...813L..31Y}. However, it remains largely unknown how to characterize these localized magnetic structures. ECM emission provides a direct measurement of the magnetic field at the emitter. The time-frequency (fine) structures in the emission may reveal distinct characteristics depending on the local magnetic field topology, a topic we will explore in this article.

\section*{Results}
\subsection*{Observations}

\begin{figure}[htbp]
	\centering
    \includegraphics[width=\linewidth]{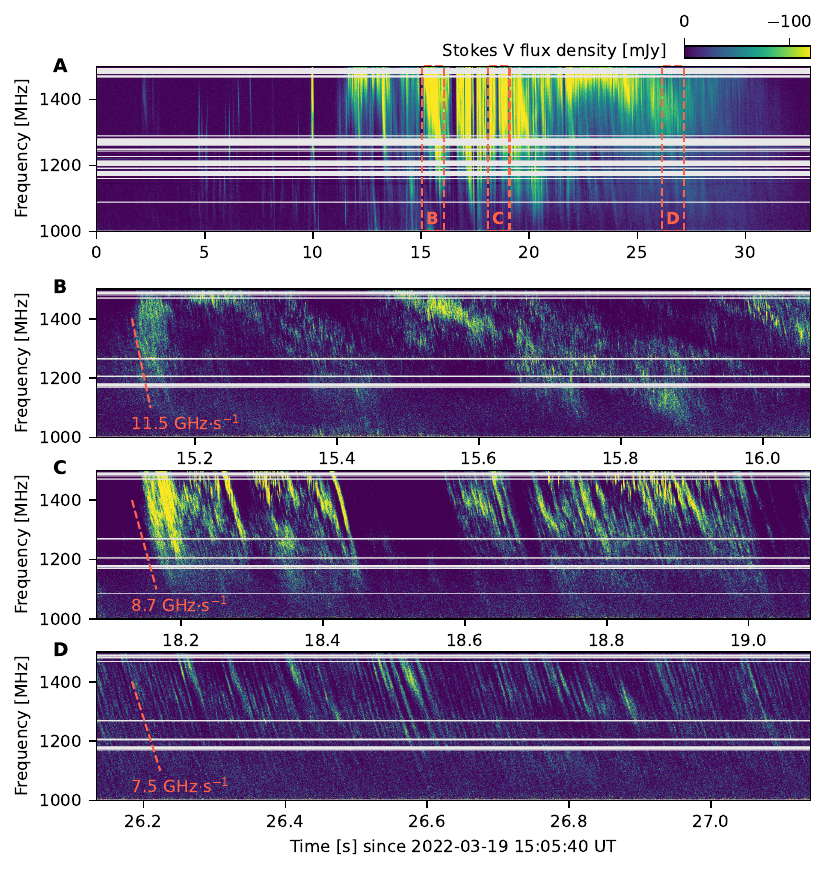}
	\caption{\textbf{Stokes V dynamic spectra of the radio bursts from AD Leo.}  (\textbf{A}) An overview of the event, presented with a resolution of 6.3 ms and 0.49 MHz. Negative values indicate right-hand circularly polarized emission, following the PSR/IEEE convention where the left-hand emission is positive and the right-hand is negative\cite{2010PASA...27..104V}. Three dashed rectangular frames denote the time ranges of the detailed fine structures shown below. Horizontal blank spaces result from radio-frequency interference (RFI) flagging. (\textbf{B to D}) Examples of the fine structures in the radio emission, presented with a resolution of 0.39 ms and 0.49 MHz. The frequency drift rates of the fine bursts are indicated by the dashed lines with their respective values.}
	\label{fig:figure1}
\end{figure}

We observed AD Leo, a nearby M3.5V flaring star, with the Five-hundred-meter Aperture Spherical radio Telescope (FAST)\cite{2011IJMPD..20..989N} at L-band (1.0-1.5 GHz) on 19 March 2022, from 14 UT to 17 UT. Throughout the three-hour observation period, only a $\sim$30 second-long radio burst event around 15:06 UT was found. The enhanced emission reaches a maximum intensity of $0.50\pm0.05$ Jy, with almost 100\% right-hand circular polarization (see Supplementary Text). As shown in the overview Stokes V dynamic spectra (Fig. \ref{fig:figure1}A), some weak bursts appear at first, then the emission intensifies from 10 s onward. The detected radio bursts span across the entire bandpass of the observation (500 MHz) and are likely to extend to higher and lower frequencies.

Owing to the extraordinary sensitivity and spectro-temporal resolution of FAST, the radio bursts are well resolved in the time-frequency plane, unveiling a wealth of fine structures (Fig. \ref{fig:figure1}, B, C, D). We found that the radio emission is highly structured, displaying many bright features with short lifetimes and narrow bandwidths (we refer to them as radio fine bursts). The fine bursts are observed across the entire frequency band and share a common drifting trend from higher to lower frequencies. Using two-dimensional (2D) fast Fourier transform (FFT) (see Materials and Methods), we determined the drift rates of the radio fine bursts as $11.5\pm4.6$ $\mathrm{GHz\;s^{-1}}$, $8.7\pm2.5$ $\mathrm{GHz\;s^{-1}}$, and $7.5\pm1.1$ $\mathrm{GHz\;s^{-1}}$ for the three examples. Notably, these drift rates represent the fastest drifting patterns ever observed in the radio emission from M dwarfs, surpassing previous reports by almost an order of magnitude (0.25-1 $\mathrm{GHz\;s^{-1}}$ in ref. \cite{1997A&A...321..841A}, $\sim$2.2 $\mathrm{GHz\;s^{-1}}$ in ref. \cite{2006ApJ...637.1016O,2008ApJ...674.1078O}, $\sim$0.45 $\mathrm{GHz\;s^{-1}}$ and $\sim$0.9 $\mathrm{GHz\;s^{-1}}$ in ref. \cite{2023ApJ...953...65Z}, all reported for AD Leo at L-band). Drift rates of the radio bursts were also measured outside the three presented examples (see Materials and Methods). Besides, we also identified a type of slower drifting patterns ($\sim$1 $\mathrm{GHz\;s^{-1}}$) in the overall structures of the fine bursts, which is most evident in Fig. \ref{fig:figure1}B. But in the current study, we will focus our attention on the ultrafast frequency drifts ($\sim$8 $\mathrm{GHz\;s^{-1}}$) of the individual fine bursts.

The radio emission we detected from AD Leo has strong circular polarization, extremely high brightness temperature ($\gtrsim 10^{17} \rm{K}$, see Supplementary Text), and short-duration radio fine structures (single-frequency duration on millisecond scale), which all converge to indicate ECM emission as the mechanism behind (see further discussion in Supplementary Text). In the non-relativistic approximation, the ECM emission frequency solely depends on the magnetic field strength, expressed as $f\approx 2.80\;n B\;\mathrm{MHz}$, where $B$ is the magnetic field strength in Gauss and $n$ is the harmonic number. Typically, only fundamental and second-harmonic emissions are relevant\cite{1985ARA&A..23..169D}, $n=1,2$.  For radio emission at $1.0-1.5$ GHz, a magnetic field strength of $357-536\;\mathrm{G}$ is required for the fundamental mode or $179-268\;\mathrm{G}$ for the second-harmonic mode. In addition, the frequency drifts of the bursts could be attributed to the intrinsic source motion, leading to a temporal variation of source magnetic field strength\cite{2008ApJ...674.1078O,2023ApJ...953...65Z} (some other contributing factors are discussed in Supplementary Text and do not substantially impact the analysis below). The observed negative frequency drifts imply that the emitting electrons are moving towards regions of weaker magnetic field. Negative drifts are also seen in Jovian "short bursts" (S-bursts)\cite{1996GeoRL..23..125Z,2014A&A...568A..53R,2023NatCo..14.5981M}, a classical type of fine structures in Jupiter's radio emissions. These bursts are emitted by electrons mirrored at the poles and propagating away from the surface towards regions with weaker magnetic field. Due to the source motion, the frequency drift rate is given by $\left|\mathrm{d}f/\mathrm{d}t\right|\approx2.80\;n\left|(\mathrm{d}B/\mathrm{d}s)(\mathrm{d}s/\mathrm{d}t)\right|=2.80nv_s\left|\mathrm{d}B/\mathrm{d}s\right|$, where $s$ is the path length of the source motion, $\mathrm{d}B/\mathrm{d}s$ is the magnetic field gradient along the path, and $v_s=\mathrm{d}s/\mathrm{d}t$ is the source velocity. As the electrons are gyrating along magnetic field lines, the magnetic field gradient is taken along the field-aligned direction. Combining the two expressions for the emission frequency and drift rate, we obtain the normalized frequency drift rate
\begin{equation}
	\left|\frac{1}{f}\frac{\mathrm{d}f}{\mathrm{d}t}\right|\approx v_s\left|\frac{1}{B}\frac{\mathrm{d}B}{\mathrm{d}s}\right|=\frac{v_s}{L_B},
\end{equation}

where $L_B=\left|B/(\mathrm{d}B/\mathrm{d}s)\right|$ is the magnetic scale height along the magnetic field line. The absolute scale height could be defined as $H_B=\left|B\right|/|\nabla B|$, and $L_B=H_B/\left|\cos\alpha\right|>H_B$ where $\alpha$ is the angle between source motion direction (also the direction of the decreasing field strength along the magnetic field line) and the magnetic gradient vector. Unless otherwise specified, magnetic scale height refers to $L_B$ in the following analysis. Taking $f=1250 \;\mathrm{MHz}$, $\left|\mathrm{d}f/\mathrm{d}t\right|=8 \;\mathrm{GHz\;s^{-1}}$, the normalized frequency drift rate ($\left|\mathrm{d}f/\mathrm{d}t\right|/f$) in our observation is $\sim6.4\;\mathrm{s^{-1}}$. For comparison, Jovian S-bursts have a typical normalized frequency drift rate of $\lesssim 1\;\mathrm{s^{-1}}$\cite{1996GeoRL..23..125Z,2014A&A...568A..53R,2023NatCo..14.5981M}, almost an order of magnitude slower than this event. The fast normalized drift rate imposes a constraint on the magnetic scale height, as the source velocity should not exceed the speed of light $c$. We infer the magnetic scale height to be less than $47\;\mathrm{Mm}$, or $0.15\;r_\star$ ($r_\star=0.44 r_\odot$ as the radius of AD Leo\cite{2015ApJ...804...64M}). According to in-situ measurements on the source regions of planetary auroral radio emission\cite{1999JGR...10410317P,2018Sci...362.2027L,2023JGRA..12831985L}, electrons with energy of kilo-electronvolts (keV) or tens of keV are typically responsible for producing the ECM emission. If assuming similar source electron energies for the AD Leo radio emission ($E\lesssim100\;\mathrm{keV}$, or $v\lesssim0.55\;c$), we may obtain a tighter constraint on the magnetic scale height, $L_B\lesssim0.08\;r_\star$.

\subsection*{Magnetic field modelling}

The magnetograms from ZDI inversion are often used to interpret the large-scale magnetic field of a star, with spatial resolution limited by the projected stellar rotational velocity, $v\sin i$. The ZDI maps of AD Leo from different epochs of spectropolarimetric measurements reveal a persistent dipole-like field topology and a nearly pole-on view over the last two decades\cite{2008MNRAS.390..567M,2018MNRAS.479.4836L,2023A&A...676A..56B}. A recent spectropolarimetric observation campaign of AD Leo between 2019 and 2020\cite{2023A&A...676A..56B} revealed a feature of decreasing axisymmetry of the large-scale field and provides a hint of an upcoming polarity reversal. The dipole component remains to account for more than $70\%$ of the total magnetic energy in the global field. There is a lower limit for the possible magnetic scale height in a pure dipole geometry, $H_B\geqslant4/(3\sqrt{17}) r_\star$ and $L_B\geqslant1/3r_\star$ (see Equations (S20) and (S21) in the Supplementary Text). Since the ZDI method reconstructs the large-scale magnetic field of AD Leo to higher orders than a dipole field, we analyze if the low magnetic scale height ($L_B<0.15\;r_\star$) region probed by the radio observation could originate from the magnetic field of AD Leo resolved by the ZDI technique.

\begin{figure}[htbp]
	\centering
    \includegraphics[width=0.8\linewidth]{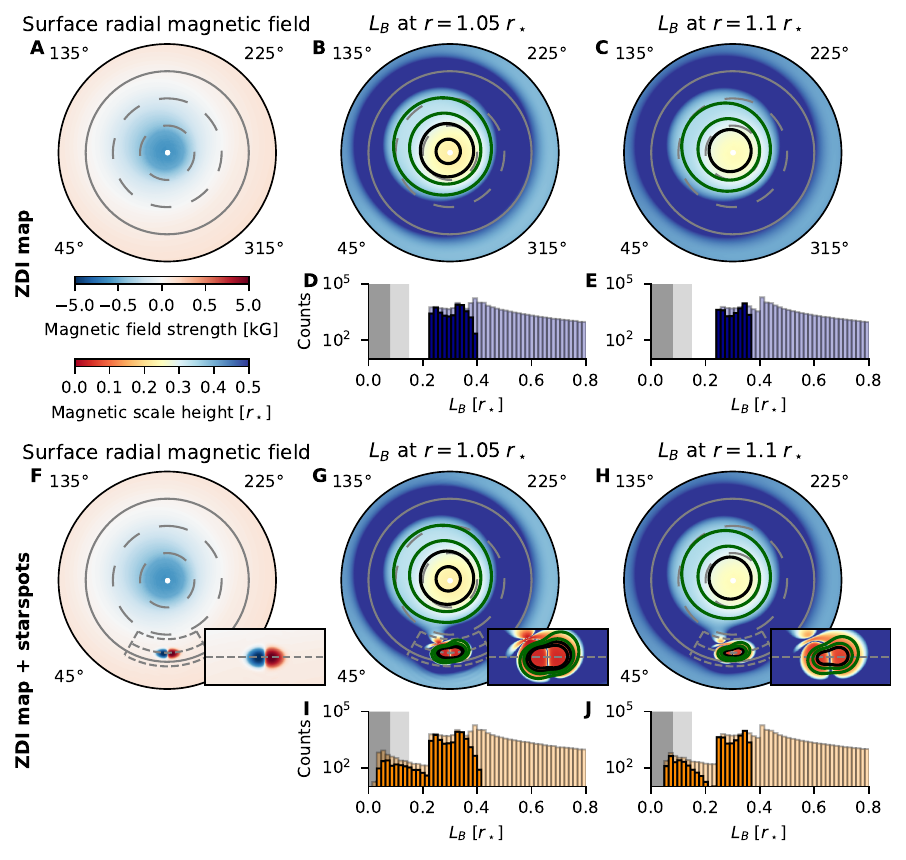}
	\caption{\textbf{Magnetic field properties of AD Leo from a flattened polar view based on two input models (A to E: ZDI map, F to J: ZDI map with a pair of starspots).} (\textbf{A}) Surface radial magnetic field ($B_r$). Blue (red) color denotes inward (outward) magnetic field. A nonlinear colorbar is used for better illustration: values between -0.5 kG -- 0.5 kG are displayed linearly in colors, while values beyond this range are shown on a logarithmic scale. The latitudes are shown by concentric circles, from outer to inner marking -30$^{\circ}$, 0$^{\circ}$, 30$^{\circ}$ (dashed), and 60$^{\circ}$ (dashed) and the longitudes (45$^{\circ}$, 135$^{\circ}$, 225$^{\circ}$, 315$^{\circ}$) are also indicated. (\textbf{B, C}) Magnetic scale height ($L_B$) at the radial distance (to the center) of $r=1.05\; r_\star$ and $r=1.1\; r_\star$. The possible source regions of the fundamental (second-harmonic) emission are enclosed by the black (green) contour lines, represented as either the gap between the two lines (i.e. the donut shapes in panel \textbf{B}) or an area encircled by the line (i.e. the round shape in panel \textbf{C}). (\textbf{D, E}) Distributions of $L_B$ from the $L_B$ maps. Light-colored histograms represent counts from all pixels in the maps and dark-colored histograms represent counts from pixels within the fundamental and second-harmonic regions. The light grey (dark grey) area marks the $L_B<0.15\; r_\star$ ($L_B\lesssim0.08\;r_\star$) threshold. (\textbf{F} to \textbf{J}) Cases of adding a pair of starspots, shown in the same format. The right-bottom sub-panels in \textbf{F} to \textbf{H} zoom in the starspots within the annulus sectors. The grey dashed lines denote the central axis of the starspots.}
	\label{fig:figure2}
\end{figure}

We employed the potential field source surface (PFSS) model\cite{1969SoPh....9..131A,2006MNRAS.370..629D,2013MNRAS.431..528J} to extrapolate the coronal magnetic field, using the surface radial field from the ZDI maps of AD Leo\cite{2023A&A...676A..56B} as the input (see Materials and Methods). We primarily based our discussions on one of the provided ZDI maps (2019b, 16 October 2019 to 12 December 2019) in the main text and analyzed three other ZDI maps (2019a, 2020a, 2020b) in the Supplementary Text which yield similar results. Using a polar-projected view, Fig. \ref{fig:figure2}A shows the radial magnetic field strength at the stellar surface and Fig. \ref{fig:figure2} (B and C) show magnetic scale height $L_B$ at two different altitudes (radial distance to the center, $r=1.05\;r_\star$ and $r=1.1\;r_\star$). We chose to plot the magnetic scale height at very low altitudes in order to find the minimum value of $L_B$, as $L_B$ in a global field geometry gradually increases with height (it will be confirmed later). At $r=1.05\;r_\star$ and $r=1.1\;r_\star$, the potential source region for fundamental ECM emission is located above $60^{\circ}$ latitude and the region for second-harmonic emission lies approximately between $30^{\circ}$ and $60^{\circ}$ latitude. $L_B$ is found to have smaller values at higher latitudes, but it does not reach values below $0.15\;r_\star$ at any location. From histograms in Fig. \ref{fig:figure2} (D and E), the $L_B$ distribution displays a minimum value lower than that of a pure dipole magnetic geometry, but still fails to extend to values lower than $0.2\;r_\star$. We conclude that the source region of the ultrafast drifting radio bursts with low $L_B$ is incompatible with the global dipole-like magnetic field extrapolated from the smooth ZDI map.

In addition to reconstructing a large-scale, smooth magnetogram, ZDI modeling also incorporates an important parameter known as the global field filling factor, $f_V$\cite{2008MNRAS.390..567M}, measured to be 9\%--16\% for AD Leo\cite{2023A&A...676A..56B}. The filling factor formalism suggests that the global magnetic field represented by the smooth ZDI map actually arises from the collective contribution of some localized regions with stronger magnetic fields, that cover only a fraction of the entire stellar surface\cite{2021A&ARv..29....1K}. Additionally, Zeeman broadening measurements on AD Leo yield an average magnetic field strength of $\sim3\;\mathrm{kG}$\cite{2014A&A...563A..35S,2017NatAs...1E.184S,2023A&A...676A..56B}, which are generally more sensitive to strong-field regions with mixed magnetic polarities. Together, these measurements suggest the presence and integral properties of small-scale magnetic structures on the surface of AD Leo, which are poorly characterized individually. One manifestation of such localized magnetism is starspots, similar to sunspots where the magnetic field is strongly concentrated on the photosphere (see further discussions on the presence of starspots on AD Leo in Supplementary Text). According to a solar observation, the magnetic scale height above a sunspot was measured to be $\sim7$ Mm ($\sim0.01\;r_\odot$) at a height of $\sim$8 Mm\cite{2006ApJ...641L..69B}, suggesting the possibility of a small magnetic scale height condition above starspots. Apart from starspots, the network field represents another possible form of local magnetic inhomogeneity at the stellar surface. However, its dominance is expected to decay substantially at coronal heights\cite{2019A&A...629A..83A}. Moreover, no ECM emission has been reported within the solar network field before. 

To understand how small-scale magnetic structures like starspots may reshape the coronal magnetic scale height distribution, we introduced a magnetic toy model that combines the ZDI map and a symmetric bipolar region representing a pair of starspots with opposite polarities\cite{2020SoPh..295..119Y} (see Materials and Methods). The key input parameters in the bipolar starspot model include the coordinate of the overall centroid of the region, the maximum strength of the magnetic field ($B_{\mathrm{max}}$) and the longitudinal distance between the centroids of the opposite polarities ($\rho$), all of which are free parameters owing to the lack of observational constraints. Initially, as shown in Fig.\ref{fig:figure2}F, the overall centroid of the bipolar region is positioned at $0^{\circ}$ in longitude and $10^{\circ}$ in latitude. $B_{\mathrm{max}}$ is set at 6 kG (twice the average surface magnetic field deduced from Zeeman broadening) and $\rho$ is set at $8^{\circ}$. Zeeman broadening analysis with a multi-component model implies that $\sim5\%$ of the stellar surface might be covered by 6 kG strength fields\cite{2023A&A...676A..56B}. The radial field of the starspots is directly added to the ZDI map. After PFSS extrapolation, we found that the bipolar magnetic region corresponds to a prominently low $L_B$ value in the upper atmosphere, while the distribution of $L_B$ outside the starspot region is merely affected (Fig. \ref{fig:figure2}, G and H). From the histograms in Fig. \ref{fig:figure2} (I and J), we spotted a distinct distribution of low $L_B$ value associated with the inserted starspots, meeting the $L_B<0.15\;r_\star$ (light grey) and $L_B\lesssim0.08\;r_\star$ (dark grey) conditions. It indicates that the starspots could lead to low $L_B$ regions consistent with the observation.

\begin{figure}[htbp]
	\centering
    \includegraphics[width=0.9\linewidth]{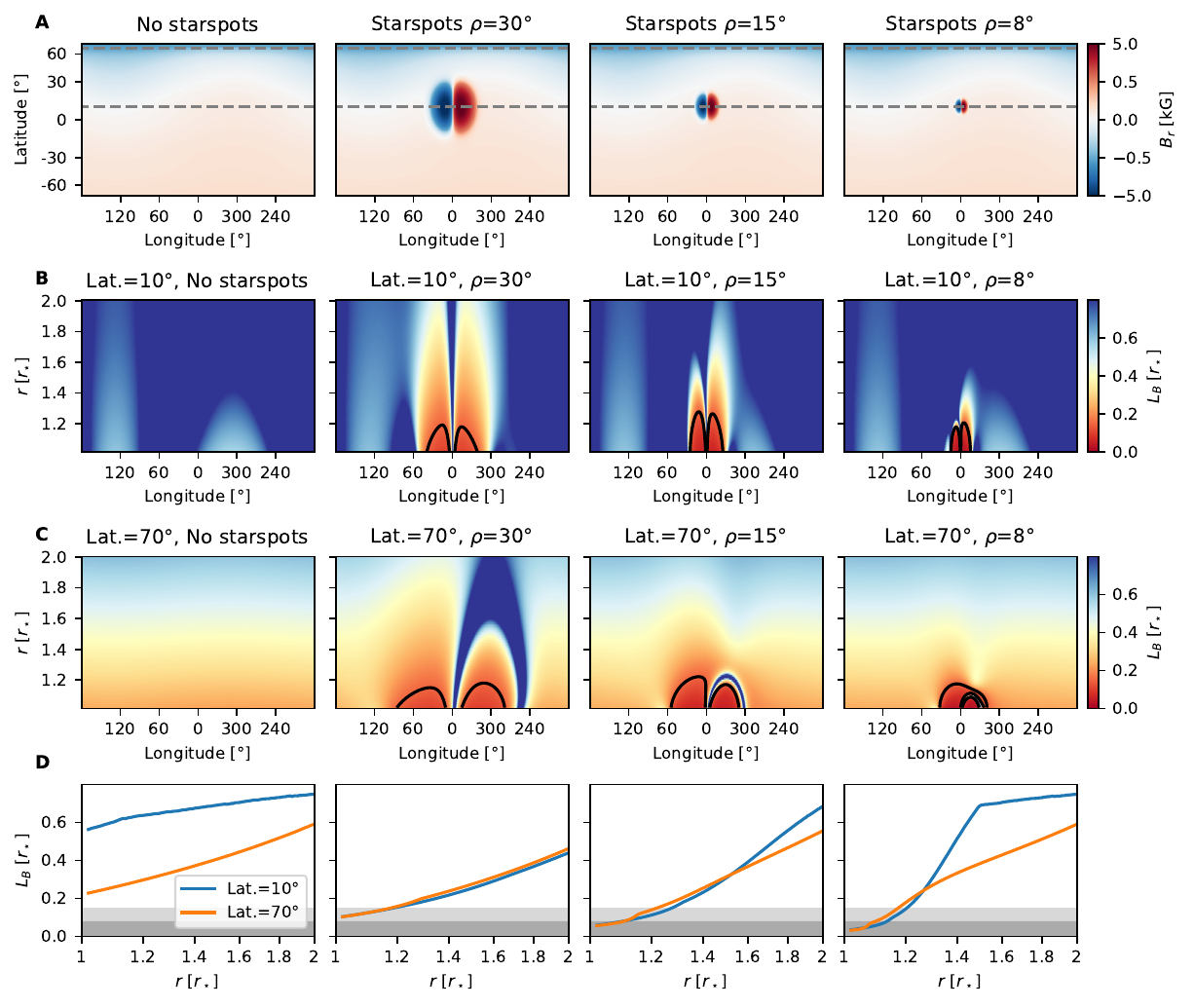}
	\caption{\textbf{The magnetic scale height distributions for adding starspots of different sizes and at different latitudes.} \textbf{(A)} Radial magnetic field incorporating the ZDI map and the starspots. From left to right are the maps without starspots and with starspots of different sizes  ($\rho=30^{\circ}, 15^{\circ}, 8^{\circ}$, $\rho$: the longitudinal distance between the centroids of the pair of starspots). The two dashed lines represent $10^{\circ}$ and $70^{\circ}$ latitudes. Starspots are initially placed at $10^{\circ}$ latitude. \textbf{(B)} The magnetic scale height ($L_B$) in a 2D slice along the central axis of the starspots ($10^{\circ}$ latitude). X-axis denotes the longitude in degree and Y-axis denotes the radial distance ($r$) in stellar radius ($r_\star$). The black curves represent $L_B=0.15\;r_\star$ contour lines. \textbf{(C)} Cases of starspots placed at $70^{\circ}$ latitude, shown in the same format as in (B). \textbf{(D)} The $L_B$ variation with radial distance. At each height, the lowest $L_B$ is selected and plotted. The blue curves represent the cases of $10^{\circ}$ latitude starspots and the orange curves represent those at $70^{\circ}$ latitude.}
	\label{fig:figure3}
\end{figure}

Next, we analyzed how the starspot properties might affect the $L_B$ distribution. We varied the latitudes (latitude=$10^{\circ}$ and $70^{\circ}$) and the scales ($\rho=30,\;15,\;8^{\circ}$) of the starspot region and plotted their $L_B$ distribution on a 2D slice along the latitudinal central axis (Fig. \ref{fig:figure3}). $B_{\mathrm{max}}$ was fixed to 6 kG in all cases (adopting different $B_{\mathrm{max}}$ values is presented in the Supplementary Text and found to only have a marginal effect on $L_B$ distributions). The first column represents the case when no starspot is inserted, serving as a comparison. The rest reveal complex morphologies in the presence of both global and local fields, with background characteristics resembling the global field and distorted features related to the local field. The variation of $L_B$ with radial distance is shown in Fig. \ref{fig:figure3}D, with the lowest $L_B$ at each height chosen and plotted here. In the absence of starspots, the magnetic scale height increases smoothly with height. Introducing local structures like starspots notably reduces the magnetic scale height, especially at lower altitudes. A smaller starspot region generally results in a smaller minimum $L_B$. As the height rises, $L_B$ increases and aligns more closely with the trend of the background global field. This is most pronounced in the case of $\rho=8^{\circ}$. Hence, the height variation of magnetic scale height reveals a transition from a region dominated by the local field to one dominated by the global field. We argue that the magnetic scale height is an important parameter to distinguish between the local field and the global field. The low magnetic scale height ($L_B<0.15\;r_\star$) inferred from the ultrafast frequency drifts unambiguously suggests that the source is above a localized strong-field structure ($\rho\lesssim30^{\circ}$), i.e., a starspot, at a very low altitude ($r\lesssim1.2\;r_\star$). Apart from the bipolar magnetic field model, we also analyzed the model of a spotted magnetogram, with the smooth surface magnetic field of the ZDI map replaced by a distribution of localized magnetic spots with a radius of $0.07\,r_\star$, covering $\sim16\%$ of the stellar surface (see the Supplementary Text). The results are also consistent with the radio constraints.

We note that a similar hypothesis was proposed in ref. \cite{2008ApJ...674.1078O} based on radio bursts from AD Leo with a drift rate of $\sim 2.2$ $ \mathrm{GHz\;s^{-1}}$. However, our discovery of ultrafast drifting bursts ($\sim8$ $\mathrm{GHz\;s^{-1}}$), combined with comparisons to the global magnetic field from ZDI measurements provides conclusive evidence. As magnetospheric-like processes (e.g. co-rotation breakdown and star-planet interaction) typically accelerates charged particles in the large-scale global field away from the surface\cite{2007P&SS...55..598Z,2012ApJ...760...59N,2013A&A...552A.119S,2017MNRAS.470.4274T}, our analysis disfavors this type of mechanisms to interpret this specific event and supports flare-associated activity powering the emission in a similar manner with the reported sunspot radio emission\cite{2024NatAs...8...50Y}. However, the lack of simultaneous optical or X-ray observations hindered us from associating the radio bursts with any flare signal which might be present in other wavelengths.

\section*{Discussion}

In this study, we present a high-time resolution radio observation on AD Leo using FAST. We detected the fastest frequency-drifting radio bursts ($\sim 8\;\mathrm{GHz}\;\mathrm{s^{-1}}$) ever observed on M-type stars, which are attributed to a source region with a very small magnetic scale height ($r<0.15\;r_\star$, or even $r\lesssim0.08\;r_\star$ for electron energies $E\lesssim100$ keV). The analysis on the coronal magnetic field suggests that the source region is incompatible with the large-scale magnetic field of AD Leo, and thus, should originate in localized small-scale magnetic structures that are currently not resolved in ZDI maps. Our argument is quantitatively verified by modeling a bipolar magnetic field region on top of the ZDI map, which reveals locally reduced magnetic scale heights consistent with the radio observation. Our investigation establishes the role of the localized magnetic structures on M dwarfs, most likely starspots, in producing some of the most intense ECM emission\cite{2008ApJ...674.1078O,2019ApJ...871..214V}. We emphasize the importance of considering localized magnetic structures and their magnetic activity when interpreting the origin of coherent radio bursts from M dwarfs in future studies.

In contrast, in a previous FAST observation campaign on AD Leo in December 2021\cite{2023ApJ...953...65Z}, fine structures with frequency drifts of a few hundred megahertz per second were reported, which could well fit in a dipole field\cite{2025A&A...695A..95Z}. The previous investigation, together with the current one, suggests that the drift rates of the radio fine structures can place a direct constraint on the magnetic gradient and magnetic scale height of the source. However, accurately measuring these magnetic field properties requires the knowledge of the source electron energy, which still seems challenging to determine. Nevertheless, we show that it becomes feasible to use drifting fine bursts to probe the multi-scale nature of the stellar magnetic field (from local structures to global topology), which promises to be an important observational constraint on the stellar magnetic field dynamo model\cite{2015ApJ...813L..31Y,2021A&ARv..29....1K}. We highlight the need for further observations on fine structures of radio bursts from other radio-active stars in order to establish this method on a larger stellar population.

Additionally, our study raises questions about the process that accelerates the electrons in the stellar active regions, which is presumably linked to coronal and chromospheric activity commonly observed in other wavelengths. There have been various competing mechanisms proposed for particle accelerations in solar flares\cite{2015Sci...350.1238C,2020Sci...367..278F,2020NatAs...4.1140C,2022Natur.606..674F,2024NatAs...8...50Y}. Their applicability to flares on M dwarfs remains unclear, especially given that the energy released in some of these flares can be orders of magnitude higher than solar ones\cite{2019MNRAS.486.1438P}. Coordinated observations across optical, ultraviolet, and X-ray bands may complement the observations of the transient radio signals from stellar active regions, providing a more complete view of plasma heating, particle accelerations, and CMEs related to stellar flares. For exoplanets orbiting active stars like AD Leo, frequent and intense magnetic activity in the stellar coronae could pose hazards to their potentially life-hosting conditions. Such investigations may help us better characterize the radiative and particle environments of the exoplanets created by their host stars, which is essential to evaluate the habitabilities of the neighboring stellar systems\cite{2022SciA....8I9743H,2022A&A...659A..10V}.


\section*{Materials and Methods}

\subsection*{Radio observations and data reduction}
We performed radio observations on 19 March 2022 with FAST, the largest single-dish radio telescope with an illuminated aperture of 300 m. The 19-beam L-band receiver mounted on FAST operates at 1.0-1.5 GHz and measures full-Stokes intensity with dual-polarization linear feeds\cite{2020RAA....20...64J}. AD Leo was continuously monitored by the central beam from 14 UT to 17 UT and data from all the 19 beams was recorded with a time resolution of 196.608 $\mu\mathrm{s}$ and a frequency resolution of 488.28 kHz. We observed the flux calibrator 3C286 10 minutes after the AD Leo observation. Throughout the observations, a series of noise diode signals about 12 K were injected into the receiving system for $\sim 1\;\mathrm{s}$ every $\sim 16\;\mathrm{s}$ and were intended for flux and polarization calibration.

We basically followed the pipeline described in detail in ref. \cite{2023ApJ...953...65Z} to carry out data reduction. The frequency-varying mismatches of the gains and phases of the two linear feeds were calibrated using the noise diode signals. The subsequent 3C286 observation was used to obtain the antenna gain of around 16 K$\cdot \mathrm{Jy^{-1}}$, which was applied to the data. The unrelated, slow background variation was determined by performing a 2\textsuperscript{nd} order polynomial fit of the data and then removed. For the overview of the event, in particular, we firstly masked the time period of the intense radio bursts and fitted the 3-minutes trend around the burst time. For the high-resolution dynamic spectra of the fine structures, we simply fitted the trend within the 1-second long period.

To flag frequency channels with severe RFIs from the dynamic spectra, we calculated the standard deviation (STD) for data at each channel. Those with comparatively larger STDs are considered corrupted by narrowband RFIs.  We determined the thresholds using the median value of the STDs from all the channels and their median absolute deviation (MAD): $\mathrm{threshold}=\mathrm{median_{STD}}+n\times\mathrm{MAD_{STD}}$. $n$ is a factor chosen with visual inspection of the dynamic spectra to retain as many usable channels as possible, ranging from 2 to 20. After obtaining the dynamic spectra from the data of the central beam, we also examined the off-beam data from the other 18 beams and identified no similar signals around the burst time (see Supplementary Text). It suggests that the results presented in this study are not caused by broadband RFIs.

As for the sign of the circular polarization, the data from four polarization products were collected as XX, YY, Re[X*Y] and Im[X*Y] for the pulsar backend of FAST\cite{2023RAA....23j4002W}. They were transformed to the Stokes parameters through I=XX+YY, Q=XX-YY, U=2$\times$Re[X*Y], and V=2$\times$Im[X*Y] in the pipeline. It means that the sign of Stokes V is left-hand circularly polarized light minus right-hand circularly polarized light, a commonly used convention (PSR/IEEE) in pulsar astronomy\cite{2010PASA...27..104V}. To ensure the consistency with the definition, we checked the data of six pulsars (B0611+22, B0919+06, B0950+08, B1133+16, B1237+25, B1933+16) with known circular polarization in the public FAST data archive. Two of them show positive sign in Stokes V (B0611+22, B0919+06), two of them show negative sign (B0950+08, B1133+16), and two of them show reversal in the sign of Stokes V in the primary component (positive to negative for B1237+25, negative to positive for B1933+16)\cite{1998MNRAS.301..235G,1998MNRAS.300..373H,1999ApJS..121..171W}. Our results of all the six pulsars are in agreement with the reported handedness in literature, implying that our convention is consistent with the one applied in pulsar astronomy.

\subsection*{Fourier transform of the dynamic spectra}

\begin{figure}[htbp]
	\centering
    \includegraphics[width=\linewidth]{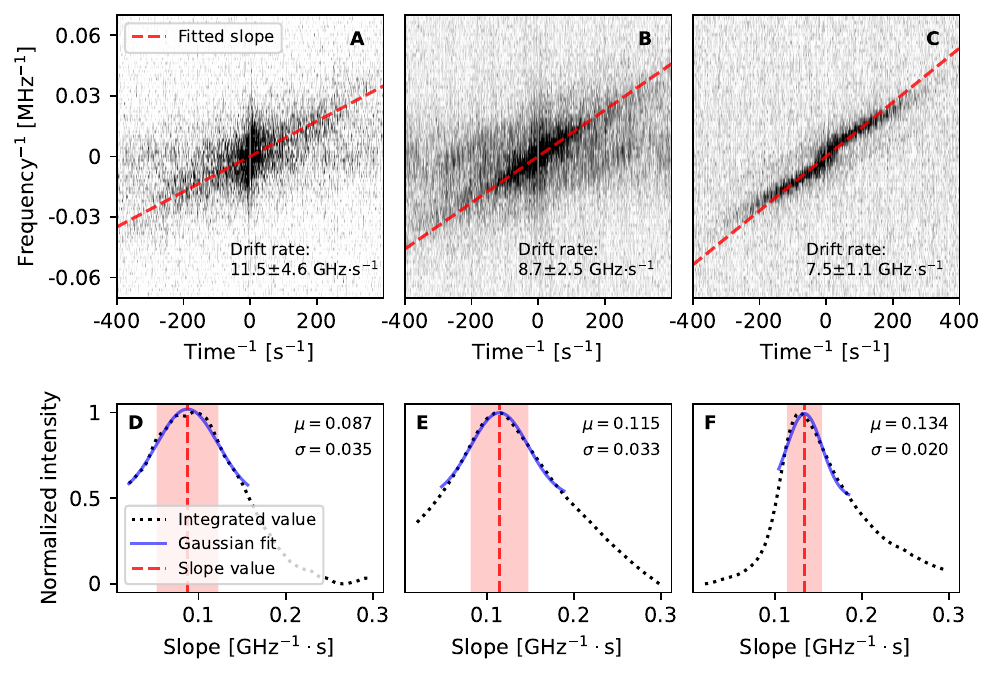}
	\caption{\textbf{Drift rate analysis using the 2D FFT diagrams of the dynamic spectra.} \textbf{(A to C)} The 2D FFT diagrams of the dynamic spectra shown in Fig. \ref{fig:figure1} (B to D). Darker color indicates stronger power. The red dashed lines denote the best-fit slopes of the central-symmetric patterns. \textbf{(D to F)} The corresponding integrated power along the lines that pass through the origin with different slopes. The black dashed curves show the normalized integrated value as a function of the slope and the blue curves show the Gaussian fit. The best-fit peak slopes and the uncertainties are shown as the red dashed lines and the shaded regions.}
	\label{fig:figure4}
\end{figure}

We used the FFT method to assess the drifting patterns in the Stokes V dynamic spectra similarly to a recent study on Jovian S-bursts\cite{2023NatCo..14.5981M}. The original data in the flagged channels with strong RFIs was first replaced by the interpolated value from adjacent valid channels, and then the dynamic spectra in $(t, f)$ plane were transformed into the Fourier plane ($t^{-1}, f^{-1}$). Here, the original drifting features are converted to perpendicular slope patterns as shown in Fig. \ref{fig:figure4} (A to C). The FFT diagrams of Fig. \ref{fig:figure4} (A and B), which correspond to the dynamic spectra of Fig. \ref{fig:figure1} (B and C), have a slope pattern that is rather diffuse, which might be related to the dispersion in the frequency drift rates of the fine bursts. The FFT diagram of Fig. \ref{fig:figure4}C, corresponding to Fig. \ref{fig:figure1}D, however, reveals a distinct shape with a kink at the center. The origin of the kink is still unknown. 

To determine the slope value (the y-coordinates over the x-coordinates, also the inverse of the drift rate), we considered a rotating line centered  in (0,0) and set it length to 800 pixels. For different rotation angles, we interpolated the power intensity of 800 points equally-distributed on the line and calculated their integral. Fig. \ref{fig:figure4} (D to F) show the normalized intensity as a function the slope value. We focused on detecting peaks in the range of slopes between $0.02-0.3 $ GHz$^{-1}\cdot$s (corresponding to the drift rates between $3-50 $ GHz$\cdot$s$^{-1}$). We specifically selected data points exceeding 0.5 intensity and applied a Gaussian function ($y=A\exp(-(x-\mu)^2/(2\sigma^2))+0.5$) to fit the trend. Fitted $\mu$ was used as the slope value and $\sigma$ was treated as the uncertainty.

\begin{figure}[htbp!]
	\centering
    \includegraphics[width=0.6\linewidth]{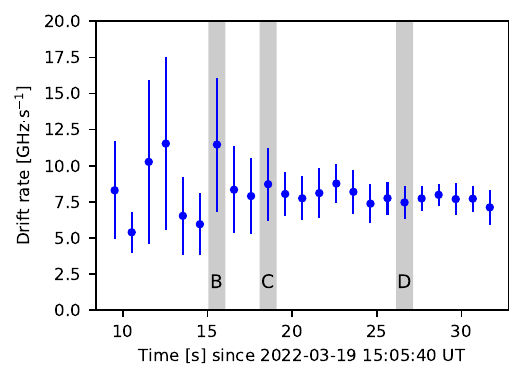}
	\caption{\textbf{The variation of the frequency drift rate of the emission fine structures at a time step of $\sim 1$ s.} The shaded regions mark the time periods of Fig. \ref{fig:figure1} (B, C, D).}
	\label{fig:figure5}
\end{figure}

We also analyzed the drift rates of the data in other time periods. The variation is displayed in Fig. \ref{fig:figure5}.

\subsection*{Bipolar magnetic field region}

Many solar active regions could be represented by a bipolar magnetic field configuration. We adopted a symmetric bipolar magnetic model from ref. \cite{2020SoPh..295..119Y}, which is based on the monitoring of solar active regions in solar cycle 24. This model has also been applied to simulate starspots\cite{2022MNRAS.509.5075S}. In this model, the centroids of the positive and negative polarities are placed at $(s_+,\phi_+)$ and $(s_-, \phi_-)$, where $s$ denotes the sine of the latitude, or cosine of the co-latitude ($s=\sin\lambda=\cos\theta$) and $\phi$ denotes the longitude. The geometrical parameters of the bipolar region, including the overall centroid ($s_0,\phi_0$), the polarity separation $\rho$, and the tilt angle with respect to the equator $\gamma$ are computed as:
\begin{equation}
	s_0=\frac{1}{2}(s_+ +s_-), \qquad \phi_0=\frac{1}{2}(\phi_+ + \phi_-)
\end{equation}
\begin{equation}
	\rho=\arccos\left[s_+s_- + \sqrt{1-s_+^2}\sqrt{1-s_-^2}\cos(\phi_+-\phi_-)\right]
\end{equation}
\begin{equation}
	\gamma=\arctan\left[\frac{\arcsin(s_+)-\arcsin(s_-)}{\sqrt{1-s_0^2}(\phi_--\phi_+)}\right]
\end{equation}

The bipolar magnetic field distribution is uniquely determined by these parameters along with the unsigned flux, $\left|\Phi\right|$. For an untilted bipole centered at $s_0=\phi_0=0$, the radial magnetic field strength could be expressed as
\begin{equation}
	B(s,\phi)=-B_0\frac{\phi}{\rho}\exp\left[-\frac{\phi^2+2\arcsin^2(s)}{(a\rho)^2}\right],
\end{equation}

where $B_0$ is a factor scaled to match the unsigned magnetic flux and $a$ is chosen to control the axial dipole component. $a=0.56$ shows a good match with the solar observations and the same value is used in our stellar model. A tilted bipolar region situated at a different ($s_0,\phi_0$) is realized by the coordinate rotation described in Appendix A in ref. \cite{2020SoPh..295..119Y} and no tilt angle is introduced in this study. From Equation (5), the maximum magnetic field strength in the region is given by $B_{\mathrm{max}}=aB_0/\sqrt{2e}\approx 0.24\;B_0$.

\subsection*{Magnetic field extrapolation}

Measuring the coronal magnetic field in both solar and stellar contexts remains a challenging task. A widely adopted approach is to use an extrapolated solution from the surface global magnetogram, known as the PFSS method\cite{1969SoPh....9..131A,2006MNRAS.370..629D,2013MNRAS.431..528J}. It assumes a current-free, potential magnetic field satisfying the boundary conditions. The inner boundary is constrained by the surface radial magnetic field and the outer boundary is known as the source surface where the field lines become completely radial. The question reduces to solving a Laplace's equation of the scalar potential $\Psi$, $\nabla^2 \Psi=0$, $\mathbf{B}^{\mathrm{pot}}=-\nabla \Psi$. In a spherical coordinate system $(r,\theta,\phi)$, the solution is written as
\begin{equation}
	\Psi=\sum_{l=1}^{N}\sum_{m=0}^{l}\left[a_{lm}r^l+b_{lm}r^{-(l+1)}\right]P_{lm}(\theta)e^{im\phi},
\end{equation}

where $r$ is scaled to the stellar radius $r_\star$, $(\theta, \phi)$ are the co-latitude and longitude, $P_{lm}$ is the associated Legendre polynomial with $l$  and $m$ denoting the degree and order of the spherical harmonic mode. The two coefficients $a_{lm}$ and $b_{lm}$ are determined by the input surface magnetogram and the source surface radius $R_{s}$. We write the radial, meridional and azimuthal components of the magnetic field as
\begin{equation}
	B_r^{\mathrm{pot}}=\sum_{l=1}^{N}\sum_{m=0}^{l}B_{lm}P_{lm}(\theta)f_l(r,R_s)r^{-(l+2)}e^{im\phi}
\end{equation}
\begin{equation}
	B_\theta^{\mathrm{pot}}=-\sum_{l=1}^{N}\sum_{m=0}^{l}B_{lm}\frac{\mathrm{d}P_{lm}(\theta)}{\mathrm{d}\theta}g_l(r,R_s)r^{-(l+2)}e^{im\phi}
\end{equation}
\begin{equation}
	B_\phi^{\mathrm{pot}}=-\sum_{l=1}^{N}\sum_{m=0}^{l}B_{lm}\frac{P_lm(\theta)}{\sin\theta}img_l(r,R_s)r^{-(l+2)}e^{im\phi},
\end{equation}

where the functions $f_l(r, R_s)$ and $g_l(r,R_s)$ are given by:
\begin{equation}
	f_l(r, R_s)=\frac{l+1+l(r/R_s)^{2l+1}}{l+1+l(1/R_s)^{2l+1}}
\end{equation}
\begin{equation}
	g_l(r, R_s)=\frac{1-(r/R_s)^{2l+1}}{l+1+l(1/R_s)^{2l+1}}.
\end{equation}

The real parts of the expressions give the actual strength of the magnetic field. The coefficients $B_{lm}$ are given by $B_{lm}=\alpha_{lm}c_{lm}$, where $\alpha_{lm}$ are the spherical harmonic coefficients of the surface radial field derived from the ZDI model and $c_{lm}$ are normalization factors,

\begin{equation}
	c_{lm}=\sqrt{\frac{2l+1}{4\pi}\frac{(l-m)!}{(l+m)!}}.
\end{equation}

The meridional and azimuthal field components on the surface are given by the combinations of the tangential poloidal field and the toroidal field, the coefficients of which are also derived independently from the ZDI inversion. However, the toroidal field is non-potential and the tangential field is no longer independent of the radial field under the source surface assumption. Therefore, we only used the information of the radial field in the ZDI model to extrapolate the magnetic field.

To determine the field topology when a pair of starspots are added on the surface, the radial component of the bipolar magnetic field is firstly decomposed to spherical harmonic terms using the following expression\cite{2012PhDT........75J,2016MNRAS.459.1533V}:
\begin{equation}
	\alpha_{lm}^{\mathrm{spot}}=(2-\delta_{m,0})\int_{\Omega} B_r^{\mathrm{spot}}c_{lm}P_{lm}(\theta)e^{-im\phi}\mathrm{d}\Omega=(2-\delta_{m,0})\int_{0}^{2\pi}\int_{0}^{\pi}B_r^{\mathrm{spot}}c_{lm}P_{lm}(\theta)e^{-im\phi}\sin\theta \mathrm{d}\theta \mathrm{d}\phi
\end{equation}

$B_r^{\mathrm{spot}}$ is the surface radial magnetic field of the starspot region. The inclusion of a factor of $(2-\delta_{m,0})$ is due to the fact that we did not consider negative $m$ values in our calculations\cite{2012PhDT........75J,2016MNRAS.459.1533V}. To balance accuracy and computation time, the spherical harmonic terms are retained to a maximum degree of $l=80$. We then added the spherical harmonic coefficients of the inserted bipole and those of the ZDI map and used Equations (7-12) to calculate the magnetic field vector. The magnetic gradient is computed by the finite difference of the magnetic field strength. The radial distance of the source surface is set at $r_s=4\;r_\star$, which is consistent with some stellar wind modelling of M dwarfs\cite{2013MNRAS.431..528J,2014MNRAS.438.1162V}.








\clearpage 

%
\bibliography{science_template} 
\bibliographystyle{sciencemag}

%
%
%
%
%
%


\section*{Acknowledgments}
This work made use of the data from FAST (Five-hundred-meter Aperture Spherical radio Telescope). FAST is a Chinese national mega-science facility, operated by National Astronomical Observatories, Chinese Academy of Sciences.
\paragraph*{Funding:}
H.T. \& J.Z. acknowledge funding by the National Natural Science Foundation of China (NSFC) grants 12425301 \& 12250006, the National Key R\&D Program of China No. 2021YFA0718600, and the Specialized Research Fund for State Key Laboratory of Solar Activity and Space Weather. H.T. also acknowledges support from the New Cornerstone Science Foundation through the XPLORER Prize. J.Z. also acknowledges support from
the China Scholarship Council (No. 202306010244). S.B. acknowledges funding by the Dutch Research Council (NWO) under the project "Exo-space weather and contemporaneous signatures of star-planet interactions" (with project number OCENW.M.22.215 of the research programme "Open Competition Domain Science- M"). T.C. acknowledges funding from NSFC grants 12273002 \& 12427804. P.Z. acknowledges funding from the ERC under the European Union’s Horizon 2020 research and innovation program (grant agreement N° 101020459—Exoradio). H.L. acknowledges funding from NSFC grant 12103004. 
\paragraph*{Author contributions:}
J.Z. scheduled the FAST observation, conducted processing and analysis of the data, implemented the magnetic field modelling and prepared the manuscript. H.T. served as the principal investigator of FAST observation project PT2021\_0019, supervised the study, organized discussions, and revised the manuscript. S.B. provided the ZDI maps. S.B. and T.C. contributed to the interpretations of the ZDI maps and magnetic field extrapolations. J.R.C., H.K.V., B.C., S.Y., P.Z., C.K.L., and Y.G. provided advice to data analysis and contributed to interpretation of the radio fine structures. FAST observations, instrument setting and monitoring were done by P.J., J.S., H.G., H.L., C.S., Z.L., and M.H.. H.L. contributed to coordinating observation and data transfer. All authors discussed the results and commented on the manuscript. 
\paragraph*{Competing interests:}
There are no competing interests to declare.
\paragraph*{Data and materials availability:}
The raw FAST observational data used in this research are available from the FAST archive (http://fast.bao.ac.cn, project ID: PT2021\_0019). The calibrated radio dynamic spectra and the results of magnetic field extrapolations are published in Zenodo at \url{https://doi.org/10.5281/zenodo.15352945}. Codes for producing dynamic spectra and performing magnetic field extrapolation can be found using the same Zenodo link, or in Github at \url{https://github.com/jiale-radioastro/ADLeo_ultrafast_drifting_bursts}. Public python packages utilized in the codes include Numpy\cite{harris2020array} (\url{https://numpy.org/}), Astropy\cite{2022ApJ...935..167A} (\url{https://www.astropy.org/}), Scipy\cite{2020SciPy-NMeth} (\url{https://scipy.org/}), and Matplotlib\cite{Hunter:2007}(\url{https://matplotlib.org/}). All other data needed to evaluate the conclusions in the paper are present in the paper and/or the Supplementary Materials.


\section*{Supplementary materials}
Supplementary Text\\
Figures S1 to S7\\
References \textit{(70-\arabic{enumiv})}\\ 

\newpage

\clearpage
\newpage


\renewcommand{\thefigure}{S\arabic{figure}}
\renewcommand{\thetable}{S\arabic{table}}
\renewcommand{\theequation}{S\arabic{equation}}
\renewcommand{\thepage}{S\arabic{page}}
\setcounter{figure}{0}
\setcounter{table}{0}
\setcounter{equation}{0}
\setcounter{page}{1} 


\begin{center}
\section*{Supplementary Materials for\\ \scititle}


        Jiale Zhang$^{1,3,4}$,
        Hui Tian$^{1,2\ast}$,
        Stefano Bellotti$^{5,6}$,
        Tianqi Cang$^{7}$,
        Joseph R. Callingham$^{3,5}$,
        Harish K. Vedantham$^{3,4}$,
        Bin Chen$^{8}$,
        Sijie Yu$^{8}$,
        Philippe Zarka$^{9,10}$,
        Corentin K. Louis$^{9}$,
        Peng Jiang$^{11,12}$,
        Hongpeng Lu$^{1,13}$,
        Yang Gao$^{14}$,
        Jinghai Sun$^{11,12}$,
        Hengqian Gan$^{11,12}$,
        Hui Li$^{11,12}$,
        Chun Sun$^{11,12}$,
        Zheng Lei$^{11,12}$,
        Menglin Huang$^{11,12}$,\\
        
    \small$^{1}$School of Earth and Space Sciences, Peking University, Beijing 100871, China.\\

    \small$^{2}$State Key Laboratory of Solar Activity and Space Weather, National Space Science Center, Chinese Academy of Sciences, Beijing 100190, China.\\
    
    \small$^{3}$ASTRON, Netherlands Institute for Radio Astronomy, Oude Hoogeveensedĳk 4, Dwingeloo, 7991 PD, The Netherlands.\\
    
    \small$^{4}$Kapteyn Astronomical Institute, University of Groningen, P.O. Box 800, 9700 AV, Groningen, The Netherlands.\\
    
    \small$^{5}$Leiden Observatory, Leiden University, PO Box 9513, 2300 RA Leiden, The Netherlands.\\
    
    \small$^{6}$Institut de Recherche en Astrophysique et Plan\'etologie, Universit\'e de Toulouse, CNRS, IRAP/UMR 5277, 14 avenue Edouard Belin, F-31400, Toulouse, France.\\
    
    \small$^{7}$School of Physics and Astronomy, Beijing Normal University, Beijing 100875, China.
    
    \small$^{8}$Center for Solar-Terrestrial Research, New Jersey Institute of Technology, Newark, NJ 07102, USA.\\
    
    \small$^{9}$LIRA, Observatoire de Paris, Université PSL, Sorbonne Université, Université Paris Cité, CY Cergy Paris Université, CNRS, 92190 Meudon, France.\\
    
    \small$^{10}$ORN, Observatoire Radioastronomique de Nan\c{c}ay, Observatoire de Paris, CNRS, Univ. PSL, Univ. Orl\'{e}ans, F-18330 Nan\c{c}ay, France.\\
    
    \small$^{11}$National Astronomical Observatories, Chinese Academy of Sciences, Beijing 100012, China.\\
    
    \small$^{12}$CAS Key Laboratory of FAST, National Astronomical Observatories, Chinese Academy of Sciences, Beijing 100101, China.\\
    
    \small$^{13}$State Key Laboratory of Public Big Data and Guizhou Radio Astronomical Observatory, Guizhou University, Guiyang 550025, China.\\
    
    \small$^{14}$School of Physics and Astronomy, Sun Yat-Sen University, Zhuhai 519082, Guangdong, China.\\
    
        \small$^\ast$Corresponding author.
    Email: huitian@pku.edu.cn\and

\end{center}

\subsubsection*{This PDF file includes:}
Supplementary Text\\
Figures S1 to S7\\

\newpage

\subsection*{Supplementary Text}

\subsubsection*{Flux density and polarization of the radio emission}

\begin{figure}[htb]
	\centering
	\includegraphics[width=\linewidth]{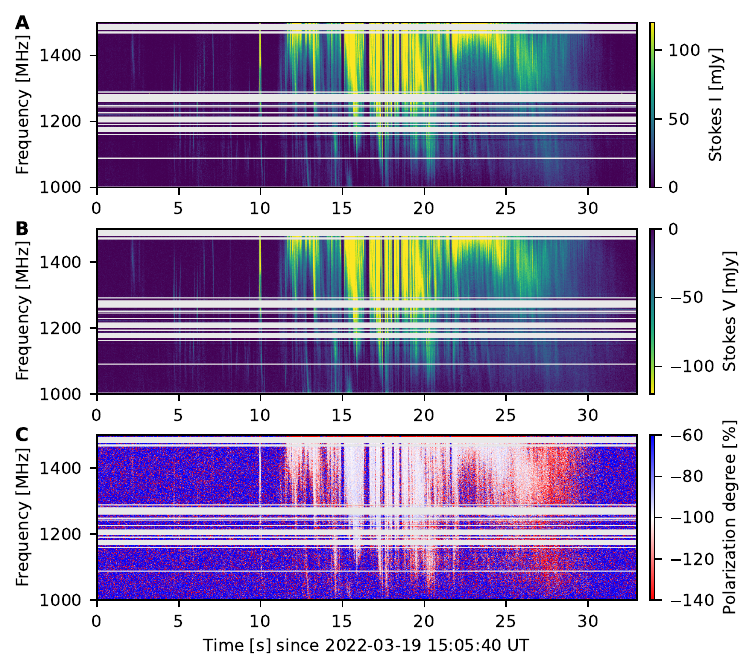}
	\caption{\textbf{Dynamic spectra of Stokes I (A), Stokes V (B) flux densities, and circular polarization degree (C).}}
	\label{fig:sfigure1}
\end{figure}

\begin{figure}[htb]
	\centering
	\includegraphics[width=0.5\linewidth]{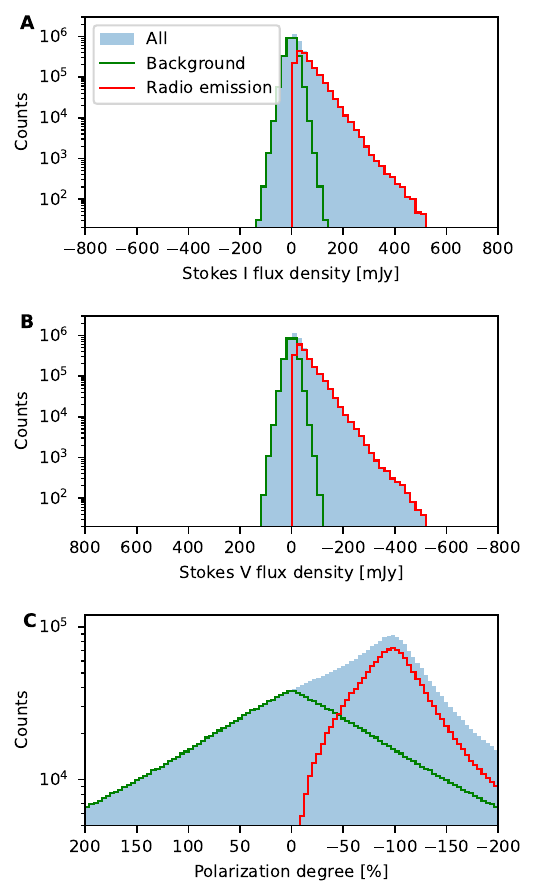}
	\caption{\textbf{Distributions of Stokes I (A), Stokes V (B) flux densities, and circular polarization degree (C) from Fig. \ref{fig:sfigure1}.} The blue filled regions indicate counts from all the pixels in the corresponding dynamic spectra. The green stairstep curves are owed to the background fluctuation and the red ones are owed to the right-hand circularly polarized radio emission.}
	\label{fig:sfigure2}
\end{figure}

We presented the dynamic spectra of Stokes I and Stokes V flux densities and circular polarization degree in Fig. \ref{fig:sfigure1}. Stokes I and Stokes V dynamic spectra display almost identical morphologies and the circular polarization degree reaches nearly 100\% (negative) in these emission patterns. As shown in Fig. \ref{fig:sfigure2}, we derived the statistical distributions of the flux density and polarization properties based on the pixels in these dynamic spectra. Partially due to the background fluctuations and measurement uncertainties, the statistics reveal broad distributions, extending to some unphysical values including negative Stokes I values and absolute circular polarization degrees $>100\%$. We assumed that the background fluctuation should adopt a symmetric distribution in positive and negative values and all the negative values in Stokes I flux density (positive values in Stokes V flux density and polarization degree) are attributed to the background. By mirroring these halves, we obtained symmetric backgrounds and subtracted them to determine the distributions of the enhanced radio emission. The maximum intensity of the emission could reach up to 0.5 Jy and the peak of the circular polarization degree is found at 98\%. We also searched for components of linear polarization of the emission but failed to find any convincing signature of degree $>2\%$. Considering the instrumental leakages, we generally conclude that the radio bursts we detected are almost 100\% circularly polarized.

\subsubsection*{AD Leo as the source of the radio bursts}

The half-power beamwidth of the FAST L-band receiver is $\sim3'$ (slightly smaller at high frequencies and larger at low frequencies)\cite{2020RAA....20...64J}, which limits the localization of the radio source to an arcminute scale. It is worth noting that there is a bright background radio source located $\sim2.3'$ away from AD Leo, which is most likely to be extragalactic\cite{1995A&A...295..123S}. However, previous AD Leo targeted radio observations have employed various techniques to rule out this extragalactic object as the source of the bursty emission, including interferometric observations\cite{1987Natur.326..678B,2019ApJ...871..214V}, and time switching scheme\cite{2006ApJ...637.1016O,2008ApJ...674.1078O}. Moreover, the extragalactic sources are known to display very low degrees of circular polarization ($<1\%$)\cite{1988ARA&A..26...93S,2023A&A...670A.124C} and could not account for the high circular polarization degree in our event. While other extragalactic radio sources may exist in the vicinity of AD Leo, their influence is also excluded for the same reason.

The known celestial objects that could be strongly circularly polarized are pulsars, main-sequence stars, brown dwarfs, and exoplanets. Pulsars are ruled out as the radio emission we detected does not show periodic pulses and lasts for only $\sim 30$ seconds. For possible stellar and sub-stellar objects other than AD Leo, we referred to the \textit{Gaia} Catalogue of Nearby Stars (GCNS)\cite{2020yCat..36490006G}, which well characterizes the objects within 100 pc of the Sun. No objects within 100 pc were found within a distance of $1.5'$ from AD Leo. Considering the radio emissivity from the stellar and sub-stellar objects, we believe that it is highly unlikely for targets beyond 100 pc to produce radio emission up to 0.5 Jy.

\begin{figure}[htb]
	\centering
	\includegraphics[width=1\linewidth]{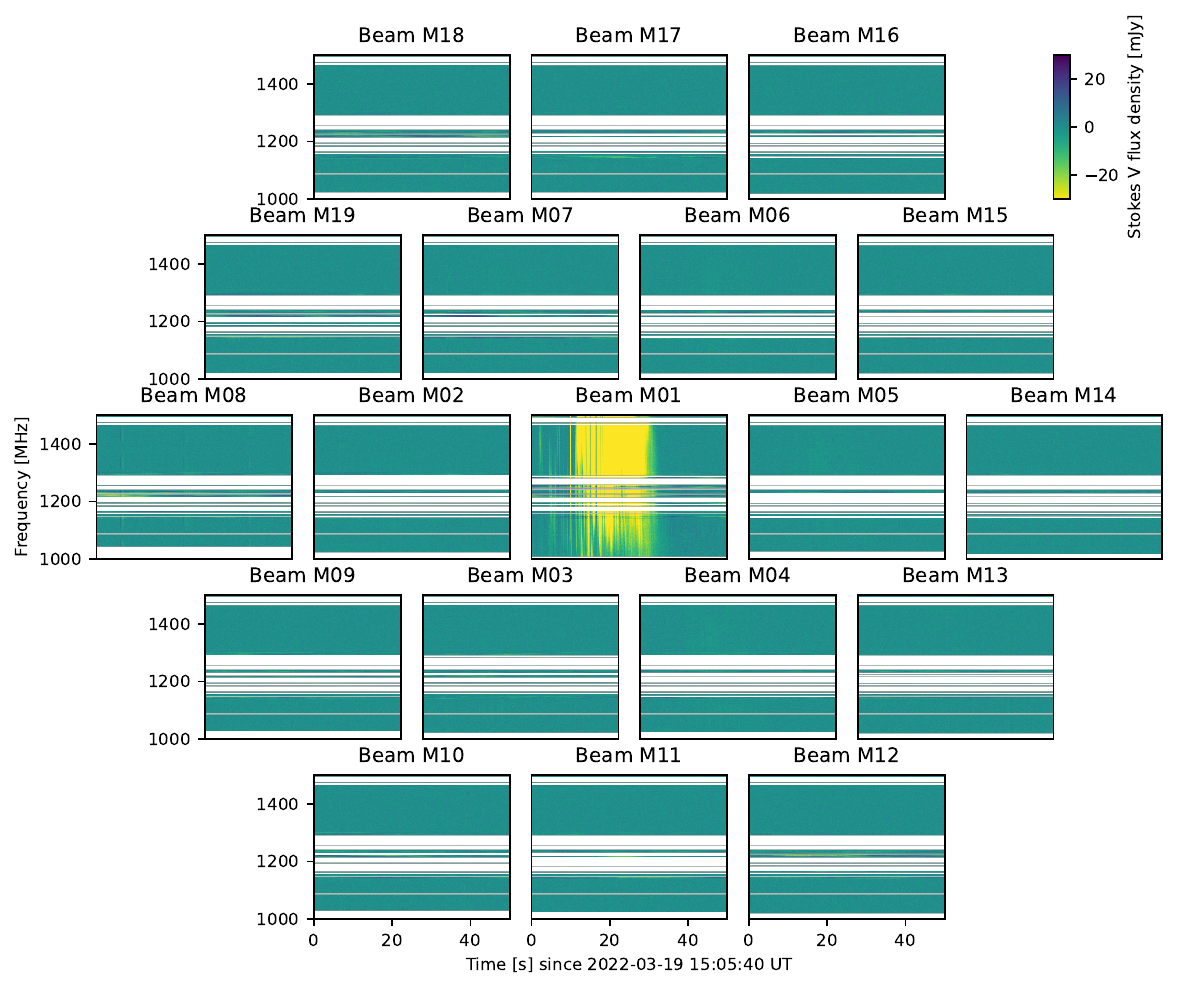}
	\caption{\textbf{Stokes V dynamic spectra of the data from all the 19 beams.} The layouts of the plots represent the relative locations of the 19 beams. No burst signal was detected from other beams during the event.}
	\label{fig:sfigure3}
\end{figure}

The possibility of the RFIs is excluded by examining the data from all the other beams (Fig. \ref{fig:sfigure3}). We also argue that the radio bursts detected in this work share many similar characteristics (intensity, polarization, fine structure patterns) with some previous radio observations of AD Leo\cite{1997A&A...321..841A,2006ApJ...637.1016O,2008ApJ...674.1078O,2023ApJ...953...65Z}, thus AD Leo is the only possible source of the emission. 

\subsubsection*{Radio emission mechanism}

There are several competing mechanisms to explain the radio emission produced in solar and stellar coronae, which are broadly categorized into incoherent and coherent mechanisms. In this context, the possibility of incoherent emission is confidently ruled out, as the radio bursts from AD Leo are almost 100\% circularly polarized and have exceptionally high brightness temperature (see below). Both the fundamental plasma emission and ECM emission could display circular polarization up to 100\%, with the former in o-mode and the latter mostly in x-mode\cite{1985ARA&A..23..169D}. The second-harmonic plasma emission tends to exhibit low polarization, often less than 20\% from solar observations\cite{1993ASSL..184.....B}. Ref. \cite{2019ApJ...871..214V} reported 22 coherent radio bursts from 5 active M dwarfs and found a consistent sense of x-mode polarization of the long-duration bursts with the predominant magnetic pole in the visible hemisphere. They attributed this consistency to ECM emission produced in a global magnetic field. For AD Leo, the visibility of the magnetic south pole throughout the rotation period implies right-hand circularly polarized emission to be o-mode. However, if the source region is embedded in some small-scale magnetic field regions with unclear magnetic polarities, the preference on polarization could be completely random. Therefore, we fail to discern the mode of polarized radio bursts based on this single event. We suggest that it would be possible to explore in future studies whether there exists a statistically dominant polarization, or a modulated pattern correlated with the rotation phase\cite{2024A&A...682A.170B} for AD Leo, which provides insights into the relationship of the emission mode and the magnetic geometry.

Next, we consider if the intensity of the radio bursts could differentiate between the two mechanisms from an efficiency perspective. We calculate the brightness temperature of the emission $T_f$ using
\begin{equation}
	T_{f}=\frac{c^2}{2k_Bf^2}\frac{S_{f}d_\star^2}{A_s},
\end{equation}

where $k_B$ is the Boltzmann constant, $f$ is the emission frequency, $S_{f}$ is the flux density, $d_\star$ is the distance of AD Leo and $A_s$ is the source size. We take $f=1250\;\mathrm{MHz}$, $S_{f}=500\;\mathrm{mJy}$, $d_\star=4.965\;\mathrm{pc}$\cite{2020yCat.1350....0G}. As for the size, the transiency of the radio fine burst (a few milliseconds at a fixed frequency) implies a considerably spatially-limited source region\cite{2008ApJ...674.1078O,2023ApJ...953...65Z}. We use the light travel time to restrain the spatial scale of the elementary radio sources, given by $l_s<c\Delta t$. $\Delta t$ is the single-frequency duration of the fine bursts, which we set at $\Delta t=5\;\mathrm{ms}$. The derived $T_f$ exceeds $10^{17}\;\mathrm{K}$ in magnitude. Such a high brightness temperature is unrealistic for plasma emission, which should saturate at the effective temperature of Langmuir turbulence ($T_L\lesssim10^{15}\;\mathrm{K}$)\cite{1985ARA&A..23..169D}. Brightness temperature up to $10^{15}\;\mathrm{K}$ is only occasionally detected for solar bursts at kilometer wavelengths, and at decameter wavelengths a temperature of $10^{10}-10^{13}\;\mathrm{K}$ is most often observed. On the other hand, the maser mechanism is believed to operate with very high efficiency, and ECM emission is often measured with extremely high brightness temperature (sometimes above $10^{20}\;\mathrm{K}$)\cite{2006A&ARv..13..229T}. Therefore, we suggest that only ECM emission could account for the high brightness temperature consistent with the intense radio bursts.

Coherent emission is usually associated with certain fine structures in the dynamic spectra, which represent an exclusive manifestation of the microscopic details in the production process. A lesson learned from the high-resolution observations on Earth and Jovian auroral emissions is that the ECM emission often displays many narrowband and short-lived discrete bursts. The most studied morphological type is the Jovian S-bursts, which have a single-frequency duration of a few milliseconds and an instantaneous bandwidth of a few kHz\cite{1982AuJPh..35..165E,2014A&A...568A..53R}. Ref. \cite{2006A&ARv..13..229T} attributed the time-frequency scale of the fine bursts to the intrinsic spatial size and appearance time of the elementary radio emitters. Radio bursts with a temporal scale down to milliseconds often imply a maser mechanism, which has been applied to interpret solar\cite{1986SoPh..104...99B} and stellar millisecond bursts\cite{2008ApJ...674.1078O,2023ApJ...953...65Z}. Plasma emission often has a duration of seconds, for instance solar type-III radio bursts\cite{2014RAA....14..773R}, which may be controlled by the conversion rate of Langmuir waves to transverse electromagnetic waves\cite{2021MNRAS.500.3898V}. Occasionally, solar plasma emission is observed with fragmented, finer features, down to subseconds or tens of milliseconds\cite{2015Sci...350.1238C,2020ApJ...897L..15M}, probably related to some inhomogeneous or turbulent conditions in the solar corona. Though ECM emission seems to be the most likely mechanism to match the millisecond time scale, it is still uncertain if plasma emission, under any possible circumstances, could power the impulsive emission that rises and decays in a few milliseconds.

Lastly, we discuss if the plasma emission could explain the fast frequency drifts in our observations. The emission frequency of the plasma emission is given by $f_{pe}=\sqrt{n_e e^2/(\pi m_e)}\approx 8.98\times 10^{3}\;\sqrt{n_e}\;\left[\mathrm{Hz}\right]$, where the electron density $n_e$, mass $m_e$ and charge $e$ are in cgs unit. Assuming plasma emission, the normalized frequency drift rate should be expressed as
\begin{equation}
	\left|\frac{1}{f}\frac{\mathrm{d}f}{\mathrm{d}t}\right|=v_s\frac{1}{2n_e}\left|\frac{\mathrm{d}n_e}{\mathrm{d}s}\right|=\frac{v_s}{2L_n},
\end{equation}

where $L_n=n_e/\left|\mathrm{d} n_e/\mathrm{d}s\right|$ is the electron density scale height. Considering the $v_s<c$ limit, we deduce a constraint of $L_n<0.06\;r_\star$. Under hydrostatic equilibrium conditions, the density scale height is given by $L_n=k_B T/(\mu m_H g)$. Here, $T$ is the coronal temperature (adopted as $T=2-10 \;\mathrm{MK}$\cite{2004ApJ...613..548M}),  $\mu$ is the mean molecular weight (adopted as $\mu=0.6$), $m_H$ is the mass of the hydrogen atom, and $g$ is the gravitational acceleration. We obtained $L_n=0.15-0.75\;r_\star$ which is much larger than the required value. Therefore, the plasma emission scenario is not consistent with the observed fast frequency-drifting patterns.

\subsubsection*{Frequency drift rate}

In this section, we evaluate factors other than source motion that may produce or affect the frequency drift. We define a time delay of $\mathrm{d}t_{obs}$ observed at a given frequency interval of $\mathrm{d}f$. $8\;\mathrm{GHz/s}$ frequency drift rate corresponds to $12.5\;\mathrm{ms}$ delay at $100\;\mathrm{MHz}$ interval. We acknowledge two types of causes of delay, the geometric delay $\mathrm{d}t_{geo}$ and dispersion delay $\mathrm{d}t_{dis}$.
\begin{equation}
	\mathrm{d}t_{obs}=\mathrm{d}t_{geo}+\mathrm{d}t_{dis}
\end{equation}

The geometric delay is due to the fact that emissions of various frequencies are produced at different locations. There are two contributions, $\mathrm{d}t_{mot}$ and $\mathrm{d}t_{pro}$. $\mathrm{d}t_{mot}$ is defined as the time difference of the source reaching different locations (source motion) and $\mathrm{d}t_{pro}$ is the time difference of light reaching the observer from different locations (light propagation).
\begin{equation}
	\mathrm{d}t_{geo}=\mathrm{d}t_{mot}+\mathrm{d}t_{pro}
\end{equation}

The dispersion delay is due to the time difference of emissions at two frequencies travelling the same distance. The contributions include the stellar corona $\mathrm{d}t_{cor}$ and the interstellar medium (ISM) $\mathrm{d}t_{ism}$.
\begin{equation}
	\mathrm{d}t_{dis}=\mathrm{d}t_{cor}+\mathrm{d}t_{ism}
\end{equation}

Therefore, we could write
\begin{equation}
	\frac{1}{\mathrm{d}f/\mathrm{d}t_{obs}}=\frac{1}{\mathrm{d}f/\mathrm{d}t_{mot}}+\frac{1}{\mathrm{d}f/\mathrm{d}t_{pro}}+\frac{1}{\mathrm{d}f/\mathrm{d}t_{cor}}+\frac{1}{\mathrm{d}f/\mathrm{d}t_{ism}}.
\end{equation}

\begin{figure}[htb]
	\centering
	\includegraphics[width=0.7\linewidth]{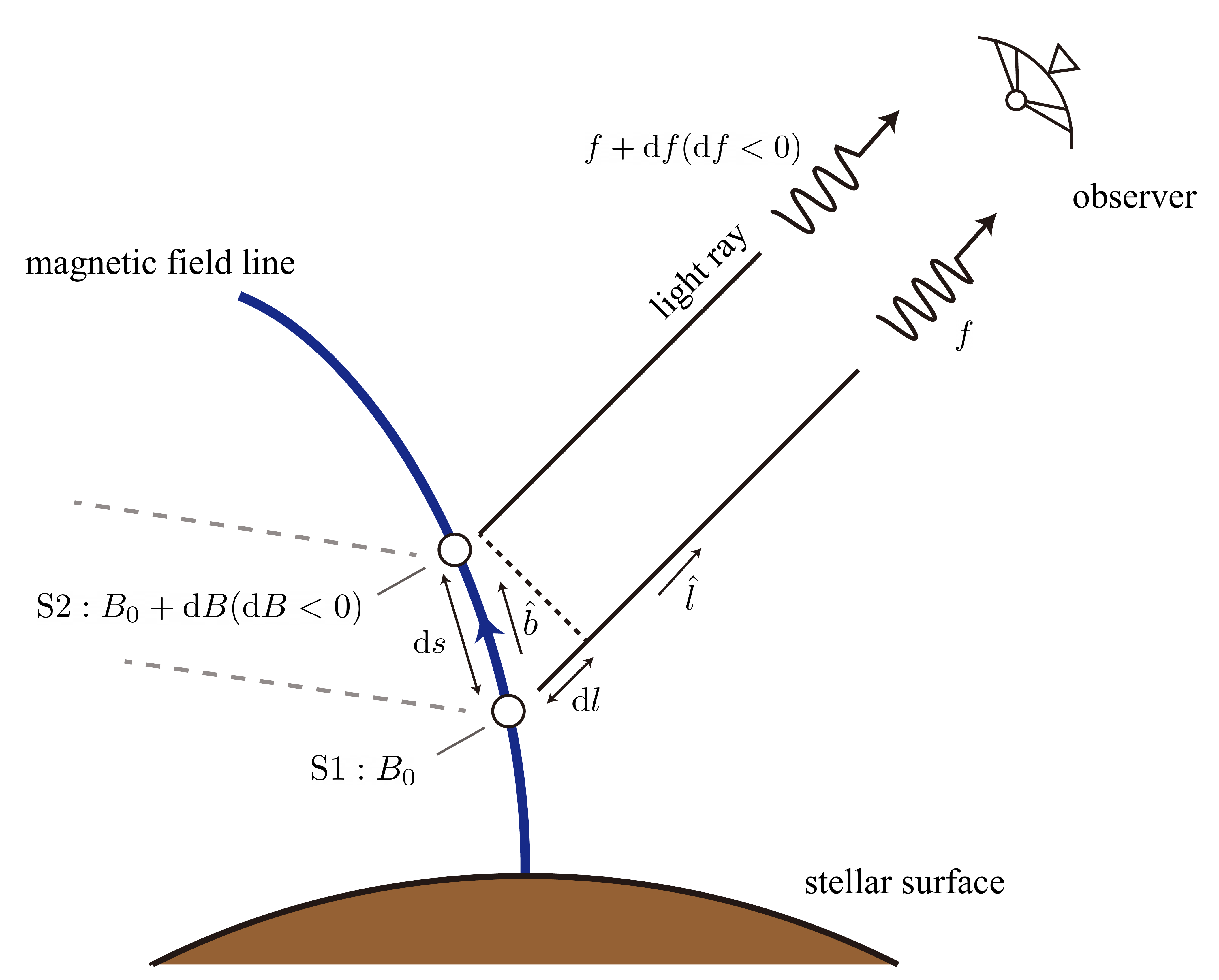}
	\caption{\textbf{A schematic diagram illustrating the geometric delay.}}
	\label{fig:sfigure4}
\end{figure}

As depicted in Fig. \ref{fig:sfigure4}, we assume that the source moves along a magnetic field line, from point S1 with magnetic field strength of $B_0$ to point S2 with strength of $B_0+\mathrm{d}B$. The distance in between is $\mathrm{d}s$ and the source velocity is $v_s$. The direction of the source motion is $\hat{b}$ and the light ray direction is $\hat{l}$.
\begin{equation}
	\mathrm{d}t_{mot}=\frac{\mathrm{d}s}{v_s}
\end{equation}
\begin{equation}
	\mathrm{d}t_{pro}=-\frac{\mathrm{d}l}{c}=-\frac{\mathrm{d}s (\hat{b}\cdot\hat{l})}{c}=-\frac{v_s}{c}(\hat{b}\cdot\hat{l})\;\mathrm{d}t_{mot}
\end{equation}

Here, $\mathrm{d}l=\mathrm{d}s (\hat{b}\cdot\hat{l})$ is the projected distance between S1 and S2 on the line of sight. It is generally accepted that the ECM emission is beamed along the narrow walls of a hollow cone almost perpendicular to the ambient magnetic field (half apex angle $\gtrsim75^{\circ}$)\cite{1998JGR...10320159Z,2006A&ARv..13..229T}. Considering electron energy less than 100 keV ($v\lesssim0.55c$)\cite{1999JGR...10410317P,2018Sci...362.2027L,2023JGRA..12831985L,2023NatCo..14.5981M}, $\left|\mathrm{d}t_{pro}\right|\lesssim0.14\;\left|\mathrm{d}t_{mot}\right|$, which does not substantially weaken the constraints discussed in the main text. However, a few planetary radio observations in the solar system have revealed possible signatures of certain ECM emissions with relatively smaller beaming angles, down to $40^{\circ}$\cite{1998JGR...10320159Z}. In our observation, a direct measurement of the ECM beaming angle is not available. Nevertheless we argue that the term $\mathrm{d}t_{pro}$ only impacts the conclusion in cases where the ECM emission has an exceptionally small beaming angle and the electron velocity approaches the speed of light, both of which are difficult to reconcile with conventional ECM theories\cite{2006A&ARv..13..229T}.

The dispersion delay is related to the frequency-dependent refractive index of the propagation medium $\mu=(1-f_{pe}^2/f^2)^{1/2}$, where $f_{pe}$ is the plasma frequency. The time difference between emissions at $f$ and $f+df$ is
\begin{equation}
	\mathrm{d}t_{dis}=\frac{1}{c}\int\left(\frac{1}{\mu(f+\mathrm{d}f)}-\frac{1}{\mu(f)}\right) \mathrm{d}l
\end{equation}

On the propagation path in the stellar corona, we assume local plasma frequency well below the emission frequency ($f \gg f_{pe}$). It is generally consistent with a supposed low-density cavity for ECM emission to escape\cite{2006A&ARv..13..229T}. In the ISM, the electron density is so low ($\sim0.1\;\mathrm{cm^{-3}}$\cite{2002astro.ph..7156C}) that the plasma frequency is extremely small compared to the emission frequency. This leads to 
\begin{equation}
	\frac{1}{\mu(f)}\approx1+\frac{f_{pe}^2}{2f^2}
\end{equation}
\begin{equation}
	\mathrm{d}t_{dis}=-\frac{e^2 \mathrm{d}f}{\pi m_e c f^3}\int n_e dl=-\frac{e^2 \mathrm{d}f}{\pi m_e c f^3} \mathrm{DM}.
\end{equation}

$\mathrm{DM}=\int n_e \mathrm{d}l$ is the dispersion measure (DM) representing the column density of the free electrons between the source and the observer. Notably, the dispersion effect inherently imposes a correlation of frequency drift rate with the emission frequency, with higher frequency corresponding to a faster drift rate. This results in a quadratic sweep in the time-frequency plane, a pronounced feature in radio pulse profiles of pulsars and fast radio bursts (FRBs)\cite{2007Sci...318..777L}. However, in our observation, the frequency drops almost linearly with time, implying that dispersion might not play the dominant role. We assume a stellar corona in a constant-gravity hydrostatic equilibrium state, where the electron density decays exponentially at a rate of density scale height $L_n$. The DM contributed by the stellar corona is given by
\begin{equation}
	\mathrm{DM}_{cor}=\int_{0}^{\infty}n_{e0}e^{-l/L_n}\mathrm{d}l=n_{e0}L_n
\end{equation}

where $n_{e0}$ is the background electron density at the height of the radio source. We first consider the possibility of an underdense, magnetized environment of the source to produce fundamental x-mode emission\cite{1984JGR....89..897M,2006A&ARv..13..229T}. We assume the local plasma frequency of the source to be $f_{pe}\approx0.1f_{ce}=0.1f$ ($f_{ce}$ as the local cyclotron frequency) and density scale height to be $L_n=(0.15-0.75)\;r_\star$ for a coronal temperature of $T=2-10\;\mathrm{MK}$. The low-density condition is hypothesized to avoid strong emission absorption at the harmonic layers of the cyclotron frequency\cite{2021MNRAS.500.3898V}. This yields $\mathrm{DM}_{cor}=0.29-1.4\;\mathrm{cm^{-3}\cdot pc}$, corresponding to a time delay of $0.12-0.61\;\mathrm{ms}$ across the $100\;\mathrm{MHz}$ interval at frequencies around $1250\;\mathrm{MHz}$. Hence, the dispersion delay is negligible compared to the observed delay. If the coronal density is more than one order of magnitude higher, the dispersion delay in the corona will become important. However, this might be a minor possibility, as the high $f_{pe}/f_{ce}$ ratio results in catastrophic gyro-harmonic absorption, preventing the emission from escaping the corona. If a high dispersion delay were to exist, it might reinforce our constraints. Since the dispersion drifts and the observed drifts are in the same (negative) direction, a longer dispersion delay would imply a shorter delay due to the source motion, requiring faster intrinsic frequency drifts (without dispersion effects) of the fine bursts. In summary, our constraints are most likely unaffected by the dispersion in the corona, or strengthened in a high-density case.

Regarding the DM of ISM,
\begin{equation}
	\mathrm{DM}_{ISM}=n_{ISM}\cdot d_\star.
\end{equation}

$n_{ISM}$ is the mean electron density of ISM, adopted as $n_{ISM}=0.1\;\mathrm{cm^{-3}}$\cite{2002astro.ph..7156C}. This gives a DM value of $0.50\;\mathrm{cm^{-3}\cdot pc}$, or a $0.21\;\mathrm{ms}$ delay across the $100\;\mathrm{MHz}$ band, which is negligible.

For the completeness of discussion, we would also consider the relativistic corrections on the ECM emission frequency. Notice that $f\approx2.80\;nB$ MHz is a first-order approximation of the ECM resonance condition, whose accurate expression is written as\cite{1979ApJ...230..621W}
\begin{equation}
	\omega-n\omega_{ce}/\gamma-k_\parallel v_{\parallel}=0
\end{equation}

where $\omega=2\pi f$ is the angular emission frequency, $\omega_{ce}=eB/m_e c$ is the (angular) cyclotron frequency of the rest mass electron, $\gamma$ is the Lorentz factor and $k_{\parallel}$ and $v_{\parallel}$ are the parallel components of the wave number and electron velocity relative to the magnetic field. The above equation is equivalent to an elliptic curve in the velocity space, known as the resonance ellipse. For electrons with energy less than a few hundred keV, "semi-relativistic case" is often used, which approximates the Lorentz factor as $\gamma^{-1}=1-v^2/(2c^2)$. The resonance curve degenerates into a circle, with its center at $v_{\parallel 0}/c=(k_\parallel c)/(n\omega_{ce})$, $v_{\perp 0}/c=0$. For shell-distribution electrons, the resonance curve is believed to center at the original point\cite{2006A&ARv..13..229T}, which means that
\begin{equation}
	\omega=n\omega_{ce}/\gamma \approx n\omega_{ce}\left(1-\frac{v^2}{2c^2}\right).
\end{equation}
 
 For loss-cone distribution electrons, ref. \cite{1982ApJ...259..844M} assumes a resonance curve being tangent to the edge of the loss cone and derives the frequency with maximum growth rate:
 \begin{equation}
 	\omega\approx n\omega_{ce}\left(1+\frac{v^2}{2c^2}\right)
 \end{equation}

The electron kinetic energy may be modulated by an electric potential drop, manifested as the perturbations of the shapes of drifting features in some Jovian S-bursts observations\cite{2009GeoRL..3614101H}. As we do not see similar signatures in our data, we assume that the electric potential drops are not present to alter the electrons' kinetic energy and Lorentz factor. Thus in both cases above, the emission frequency is still approximately proportional to the cyclotron frequency, or the magnetic field strength. Relationship of $(d\omega/dt)/\omega=(dB/dt)/B$ holds and the magnetic scale height calculation is unaffected. This is valid for electrons with energy less than 100 keV, which is thought typical for producing ECM emission\cite{1999JGR...10410317P,2018Sci...362.2027L,2023JGRA..12831985L,2023NatCo..14.5981M}.

Electrons with much higher energy, reaching up to hundreds of keV or a few MeV, are usually responsible for synchrotron emission due to stronger relativistic effects. It is still uncertain from an observational perspective if strongly relativistic electrons could generate ECM emission, but there are some simulation attempts considering the fully relativistic conditions of ECM emission from relativistic electrons and found higher growth rates compared with semi-relativistic conditions\cite{2023ApJ...944...37Z}. In this extreme case, semi-relativistic approximation may not be applicable and we need to rely on kinetic particle simulations for general solutions, which is beyond the scope of the current work. However, we have found at least one clue that probably diminishes the possibility. Relativistic and ultra-relativistic electrons are known to induce notable broadening of the emission bandwidth. In ref. \cite{2023ApJ...944...37Z}, the emission bandwidth expands from 1\% to 10\% of the cyclotron frequency for electron energy from 20 keV to 200 keV. In our observation, the instantaneous bandwidth of the individual fine burst is very narrow, reaching down to $\triangle f/f \sim 1\%$. This implies that the source electrons may not extend to such a high energy level.

\subsubsection*{Alternative magnetic field models}

In this section, we will discuss some alternative magnetic field models that are not fully analyzed in the main text.

1. Dipole magnetic field geometry

In a spherical coordinate system $(r,\theta,\phi)$ where the z-axis is along the magnetic axis, the magnetic field of a magnetic dipole is expressed as

\begin{equation}
	\mathbf{B}=B_r \hat{r}+B_\theta\hat{\theta}+B_\phi\hat{\phi}=\frac{B_p r_\star^3}{r^3}\cos\theta \hat{r}+ \frac{B_p r_\star^3}{2r^3}\sin\theta \hat{\theta},
\end{equation}

where $B_p$ is the magnetic field strength at the magnetic poles and $r_\star$ is the stellar radius. The magnetic field strength is given by
\begin{equation}
	B=\frac{B_p r_\star^3}{2r^3}\sqrt{1+3\cos^2\theta},
\end{equation}

and the magnetic field gradient is given by
\begin{equation}
	\nabla B=-\frac{3B_p r_\star^3}{2r^4}(\sqrt{1+3\cos^2\theta}\;\hat{r}+\frac{\sin\theta\cos\theta}{\sqrt{1+3\cos^2\theta}}\;\hat{\theta})
\end{equation}

It is proved that the absolute magnetic scale height $H_B$ has a minimum value of
\begin{equation}
	H_B=\frac{B}{\left|\nabla B\right|}\geqslant\frac{4}{3\sqrt{17}}r\geqslant\frac{4}{3\sqrt{17}}r_\star\approx0.323\,r_\star
\end{equation}

which is obtained when $r=r_\star$ and $\theta=\arccos(\frac{\sqrt{5}}{5})$. For the magnetic scale height along the field lines $L_B$, the minimum value is found at
\begin{equation}
	L_B=\frac{B}{\left|\nabla B\cdot\hat{b}\right|}\geqslant\frac{1}{3}r\geqslant\frac{1}{3}r_\star\approx0.333\,r_\star
\end{equation}

when $r=r_\star$ and $\theta=0^{\circ}$ or  $180^{\circ}$. $\hat{b}=\mathbf{B}/B$ is the magnetic unit vector. Hence it is impossible to have $L_B<0.15 \,r_\star$ in a purely dipolar magnetic field topology.

2. ZDI maps of AD Leo between 2019 and 2020

There is increasing evidence that suggests long-term variations of the large-scale magnetic field of M dwarfs, potentially implying magnetic cycles \cite{2023A&A...676A..56B}. For instance, ref. \cite{2018MNRAS.479.4836L} compared the magnetic field maps of AD Leo in 2012 and 2016 and reported a substantial decline in the total magnetic energy and stronger concentration at local fields between these four years. In Ref. \cite{2023A&A...676A..56B}, a more time-continuous magnetic monitoring campaign was carried out between 2019 and 2020, revealing a decreasing field axisymmetry and an increasing dipole obliquity. They attributed these features as a hint of an upcoming polarity reversal of AD Leo. 

\begin{figure}[htbp!]
    \centering
    \includegraphics[width=0.9\linewidth]{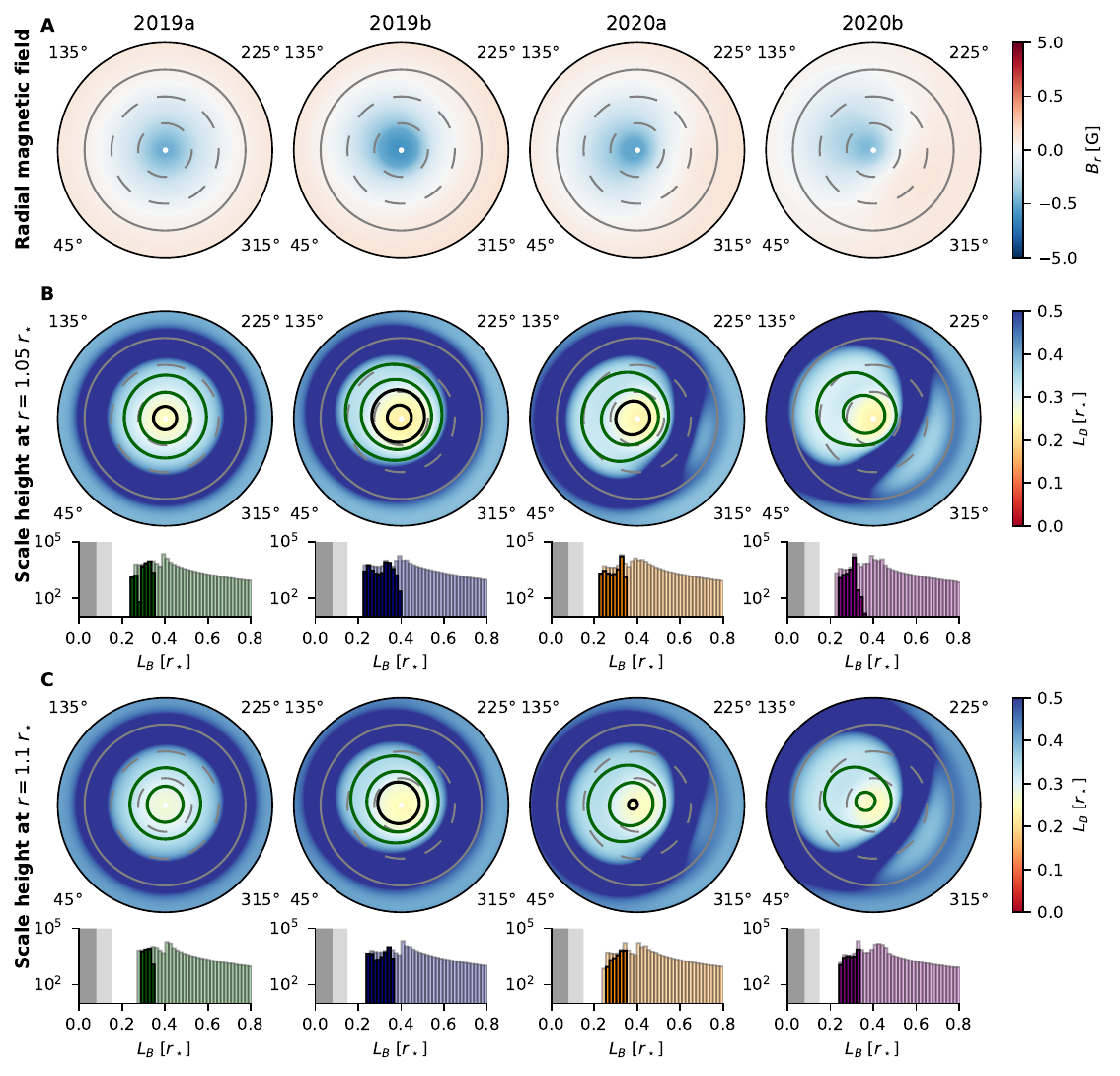}
    \caption{\textbf{Magnetic field properties of AD Leo extrapolated from four ZDI maps (2019a, 2019b, 2020a, 2020b from left to right) in ref.\cite{2023A&A...676A..56B}}. \textbf{(A)} The radial magnetic field at the surface; \textbf{(B)} The magnetic scale height at $r=1.05 \;r_\star$ and the corresponding distributions. The formats are the same with Fig. \ref{fig:figure2}(B,D) in the main text. \textbf{(C)} The magnetic scale height at $r=1.1 \;r_\star$ and the distributions.}
    \label{fig:sfigure5}
\end{figure}

Here we discuss if the evolution of the global field may substantially affect the robustness of the magnetic scale height analysis. As shown in Fig. \ref{fig:sfigure5}, we did the same $L_B$ analysis on four ZDI maps of 2019a (15 April 2019 to 21 June 2019), 2019b (16 October 2019 to 12 December 2019), 2020a (26 January 2020 to 12 March 2020) and 2020b (8 May 2020 to 10 June 2020) obtained from the near-infrared observations between 2019 and 2020 \cite{2023A&A...676A..56B}. The four maps have a corresponding tilt angle of the magnetic axis of $2.5^{\circ}$, $12.5^{\circ}$, $19.5^{\circ}$ and $38.0^{\circ}$, respectively. We found that the magnetic scale height maps at $r=1.05\;r_\star$ and $r=1.1\;r_\star$ share a similar topological variation with the surface field maps. The statistical distributions of $L_B$ at these heights subtly vary and all stay above $0.2\;r_\star$. Therefore, although the global magnetic field of AD Leo may evolve over time, our conclusion remains solid that a low magnetic scale height region could not originate from global-scale magnetic structures. 

3. The spotted magnetogram of AD Leo considering the filling factor

The current ZDI model adopts a filling factor formalism, implying that only a fraction of the stellar surface contributes to the observed large-scale magnetic field. One possible way to incorporate the filling factor into the ZDI map is by replacing the smooth magnetic field with a distribution of semi-randomly placed magnetic spots, which covers a fraction of the stellar surface and follows the large-scale field configuration\cite{2021A&ARv..29....1K}.

\begin{figure}[htbp!]
    \centering
    \includegraphics[width=\linewidth]{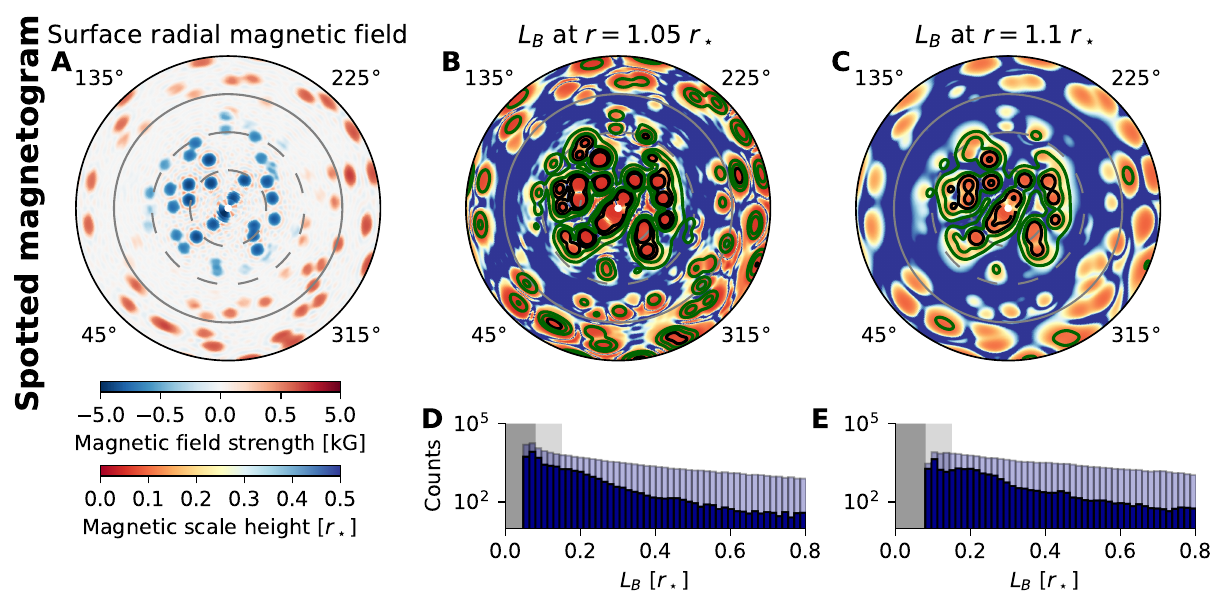}
    \caption{\textbf{Magnetic field properties of a spotted magnetogram model of AD Leo, considering the filling factor of $\sim$16\%. The formats are the same with Fig. \ref{fig:figure2}(A-E) in the main text.}}
    \label{fig:sfigure6}
\end{figure}

To explore this possibility, we divided the smooth ZDI map of AD Leo (2019b) into $8\times8$ equal-sized segments. Within each segment, we replaced the smooth magnetic field with two randomly placed circular magnetic spots, each with a uniform radius of 0.07 $r_\star$. The magnetic field strength of the spots is set to conserve the magnetic flux of each segment. Altogether, these magnetic spots cover $\sim$16\% of the entire stellar surface, in agreement with the reported filling factor \cite{2023A&A...676A..56B}. We then used Equation (13) to decompose the radial magnetic field to spherical harmonic terms and extrapolated the magnetic field topology. The surface radial magnetic field and the magnetic scale height distributions at two radial distances (1.05 $r_\star$ and 1.1 $r_\star$) are shown in Fig. \ref{fig:sfigure6}. We find that the spotted magnetogram model could satisfy the constraints from radio observations, with regions of low $L_B$ spatially coinciding with the locations of the introduced magnetic spots.

4. Bipolar magnetic field regions with different maximum magnetic field strengths

In the main text, we show that the magnetic scale height above the bipolar magnetic region is primarily controlled by the size of the region. For the completeness of discussion, we analyze here if the magnetic field strength at the magnetic spots could also be an important contributing factor. 

\begin{figure}[htbp!]
    \centering
    \includegraphics[width=\linewidth]{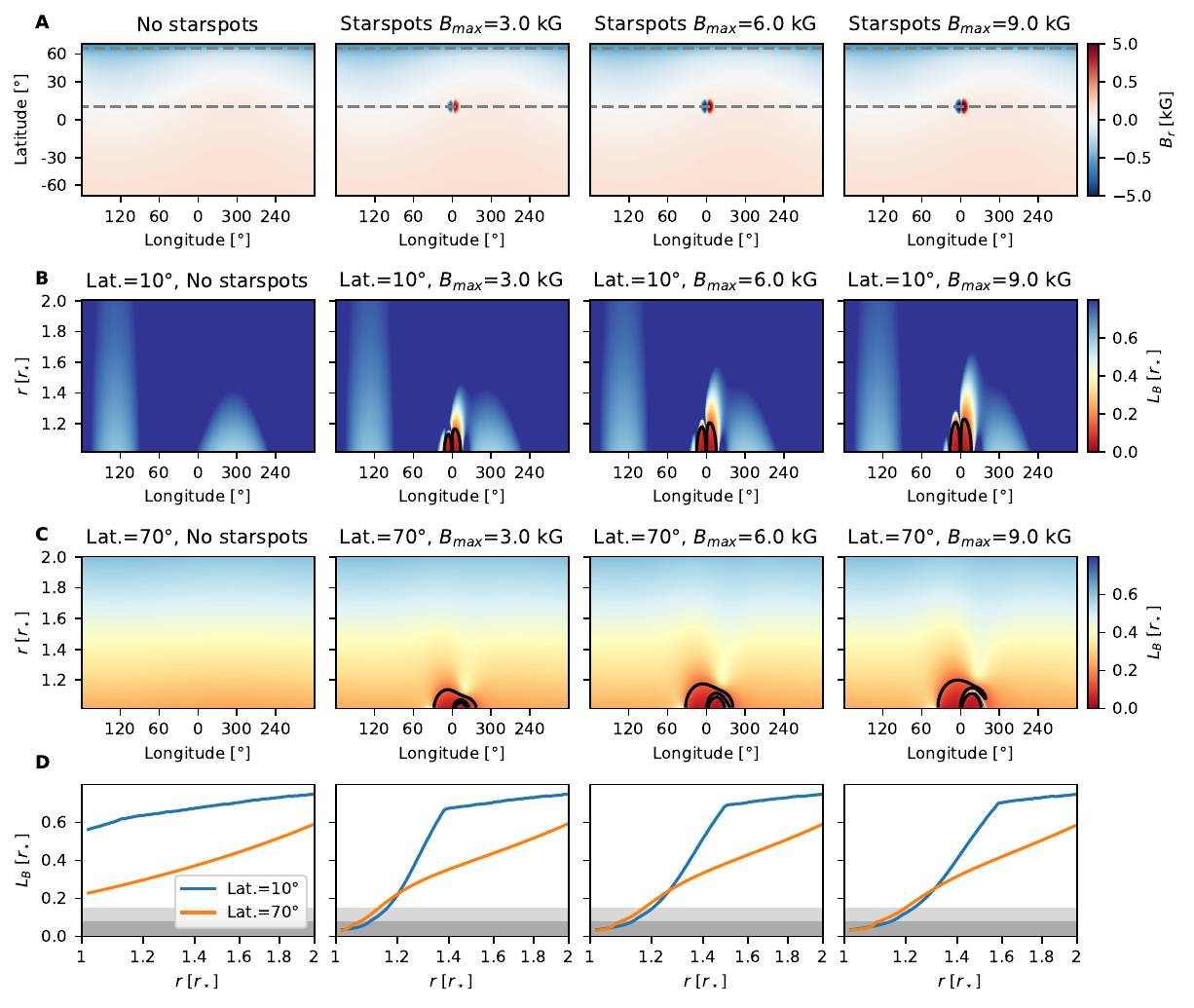}
    \caption{\textbf{The magnetic scale height distributions for adding starspots with different magnetic field strengths and at different latitudes.} \textbf{(A)} Radial magnetic field incorporating the ZDI map and starspots. From left to right are the cases without starspots and with starspots having different magnetic field strengths ($B_{\mathrm{max}}=3, 6, 9\;\mathrm{kG}$, $B_{\mathrm{max}}$: the maximum magnetic field strength). \textbf{(B,C)} The magnetic scale height ($L_B$) in a 2D slice along the central axis of the starspots (10$^{\circ}$ and $70^{\circ}$ latitudes, respectively). \textbf{(D)} The $L_B$ variation with radial distance. The other formats are the same with Figure \ref{fig:figure3} in the main text.}
    \label{fig:sfigure7}
\end{figure}

We fixed the size of the magnetic bipolar region ($\rho=8^{\circ}$) and varied the maximum magnetic field strength ($B_{\mathrm{max}}=3\;\mathrm{kG},\;6\;\mathrm{kG},9\;\mathrm{kG}$) and the latitudinal position ($10^{\circ}$ and $70^{\circ}$). The results are presented in Fig. \ref{fig:sfigure7}. We found that $B_{\mathrm{max}}$ has a very marginal effect on the $L_B$ distributions. But on the other hand, $B_{\mathrm{max}}$ changes the local magnetic field strength and thus have an important influence on the emission frequency produced at different heights.

\subsubsection*{Presence of starspots on AD Leo}

Solar magnetic field on the surface (photosphere) is highly inhomogeneous. The strongest solar magnetic fields are observed in sunspots, often reaching kilogauss strengths. In contrast, the disk-averaged unsigned solar magnetic flux is only a few Gauss\cite{2022ApJ...935....6H}. Sunspots typically appear in pairs or groups, forming active regions which are the primary sources of solar eruptive activities. It is expected that similar small-scale magnetic structures also exist on the surface of M dwarfs\cite{2015ApJ...813L..31Y,2015A&A...573A..68Y,2016ApJ...833L..28Y}. For instance, using magneto-hydrodynamic simulations, ref. \cite{2015ApJ...813L..31Y} modeled a low-mass fully-convective star with a rotation period of 20 days, revealing a complex interplay of a global field and numerous localized bipolar fields. The study found that the ZDI map could only recover a dipole-dominated large-scale field with an average strength of 450 G, which accounts for 20\% of the total magnetic flux across all spatial scales on the surface. This is generally consistent with the known discrepancies between the average field measurements from Zeeman broadening and ZDI\cite{2009A&A...496..787R}. Though the simulation is set on a fully-convective star, similar results could be extrapolated to other low-mass stars, including AD Leo (with a mass of 0.42 $M_\odot$\cite{2023MNRAS.522.1342C}), which lies slightly above the theoretical fully-convective boundary at 0.35 $M_\odot$\cite{1997A&A...327.1039C}. ZDI maps of AD Leo typically yield a mean magnetic field strength of $\sim$ 100 G, much smaller than the $\sim$3 kG measurement from Zeeman broadening\cite{2023A&A...676A..56B}. This strongly suggests a distribution of kilogauss strength local magnetic fields of mixed polarities on the surface that are currently not resolved. Besides, the global field filling factor incorporated in the ZDI model indicates that the large-scale magnetic field originates from a global configuration of small-scale fields \cite{2021A&ARv..29....1K}, although this aspect is often overlooked when interpreting ZDI maps.

There are also other observational evidences to support the presence of starspots on AD Leo. Photometric observations have revealed a flux modulation of AD Leo consistent with the rotation period\cite{2012PASP..124..545H,2022A&A...667L...9S,2023PASP..135f4201B,2020PASJ...72...68N}. This phenomenon is commonly explained by one (or several) starspot rotating into and out of view. Besides, rotational modulation on the radial velocity of AD Leo has been discovered. It is initially attributed to an exoplanet in spin-orbit resonance\cite{2018AJ....155..192T} but later re-evaluated as the possible influence from starspots and their activities\cite{2020A&A...638A...5C,2022A&A...666A.143K,2023A&A...674A.110C}. Furthermore, as a very active star, AD Leo hosts frequent flaring activities which are essentially powered by the evolution of magnetic field in the stellar active regions. However, the observational constraints on the properties of AD Leo starspots (including locations, sizes, polarities) remain scarce. The most common methods to trace and characterize starspots include Doppler imaging, spot model inversion on the light curves, and direct imaging (see ref. \cite{2009A&ARv..17..251S} for a review). Application of Doppler imaging is typically limited to some fast-rotating stars (projected rotational velocities $v\sin i \gtrsim$ 20 km s$^{-1}$)\cite{2005MNRAS.359..711M}, whereas AD Leo has a rotation speed of $v\sin i =$ 3 km s$^{-1}$\cite{2023A&A...676A..56B}. Direct imaging via long-baseline optical/near-infrared interferometry is currently available for only a few giant stars\cite{2016Natur.533..217R,2021ApJ...913...54P}. Spot model inversion based on photometric observations might be the optimal method to reconstruct the spatial distribution of starspots on AD Leo\cite{1998ARep...42..649A,2017ARep...61..221A,2022A&A...667L...9S}. Ref. \cite{2017ARep...61..221A} found that the starspots of AD Leo are located at low-latitude regions (0-10$^{\circ}$) and cover 10\%--30\% of the entire surface. Limitations of the spot model inversion mainly come from the degeneracy of the spot properties. Especially for AD Leo with a nearly pole-on geometry, the inversion will bias towards recovering starspots at low latitudes which could modulate the light curves. It is not clear if there might be starspots in the polar region, where the large-scale magnetic field is apparently stronger.
 
Given the limited characterization of starspots on AD Leo, it is not feasible to construct a magnetic field model that faithfully represents reality. Therefore, we consider introducing a magnetic field structure that is not in conflict with current magnetic field measurements. This requires a region whose size is unresolved by the ZDI technique and whose field strength agrees in magnitude with those inferred from Zeeman broadening measurements. Considering the low rotation velocity $v\sin i$ of AD Leo, spherical harmonic expansion was retained to a maximum degree of $l_{\mathrm{max}}=8$ in the ZDI model\cite{2023A&A...676A..56B}, setting the limit on the spatial resolution of the reconstructed magnetogram. The longitudinal separations between the two magnetic poles ($\rho\leqslant30^{\circ}$) in our model ensures that the introduced structures could not be resolved. Meanwhile, the Zeeman broadening analysis gives clues on the possible magnetic field strengths on the surface, ranging from 1 kG to 10 kG with corresponding filling factors\cite{2023A&A...676A..56B}. The field strength of the starspots we introduce to the model ($B_{\max}=6$ kG) is within the scope. Though we have only explored limited examples of possible starspot fields in this study, our conclusions are valid to a much broader range of configurations. We hope that future observations may provide more useful information on the starspot properties and morphologies.

\end{document}